# Dynamic Risk in the U.S. Banking System: An Analysis of Sentiment, Policy Shocks, and Spillover Effects

Haibo Wang[1]
A. R. Sánchez Jr. School of Business, Texas A&M International University, Laredo, Texas, USA

Jun Huang
Department of Management and Marketing, Angelo State University, San Angelo, Texas, USA

Lutfu S. Sua
Department of Management and Marketing, Southern University and A&M College, Baton Rouge, Louisiana, USA

Jaime Ortiz
Department of International Business and Entrepreneurship, The University of Texas Rio Grande Valley, Edinburg, Texas, USA

Jinshyang Roan
Department of Business Administration, Soochow University, Taipei City, Taiwan

Bahram Alidaee
Department of Marketing, School of Business Administration, The University of Mississippi, Oxford, Mississippi, USA

**Abstract**
This study analyzes the high-frequency daily returns and volatility spillover dynamics for eight U.S. banks from March 18, 2022, to March 15, 2023. We examine the interrelationships between bank stock returns and three economy-wide variables: market sentiment (VIX), monetary policy (SOFR), and economic policy uncertainty (EPU). We integrate correlation, cointegration, and connectedness (3C) analyses to capture dynamic systemic risk transmission using a time-varying parameter vector autoregression (TVP-VAR) model with 30-day rolling windows. Results show that spillovers are episodic and asymmetric. A subset of banks (SIVB, FRC, WAL) and SOFR act as key net transmitters of shocks, while other banks, VIX, and EPU predominantly absorb spillovers. These roles intensify during stress periods, especially May–July 2022. Our findings confirm persistent systemic risks within the banking network and highlight the importance of real-time monitoring to strengthen financial stability.

**Keywords:** Market volatility spillovers; dynamic conditional connectedness; time-varying parameter vector autoregressions; monetary policy; uncertainty

---

[1] Address correspondence to Haibo Wang, Ph.D., Division of International Business and Technology Studies, A. R. Sanchez, Jr. School of Business, Texas A&M International University, 5201 University Boulevard, Laredo, Texas, USA. Email: hwang@tamiu.edu

1. **Introduction**

Bank crises are recurrent events driven by various factors and whose impacts vary in scale and by region. For instance, the 1907 banking panic was the first worldwide crisis in the 20$^{th}$ century. Widespread public distrust in the banking system led to a bank run that eventually collapsed the stock market. In response, the U.S. Congress established the Federal Reserve System in 1913 to mitigate the impact of future banking crises. However, it failed to prevent the banking episodes of 1930 and 1931, when investors directed their savings or borrowed funds into overvalued stocks, resulting in another stock market crash. To address the aftermath, the Federal Deposit Insurance Corporation (FDIC) was created in 1933 to ameliorate the damage.

The global-scale banking crisis during 2007-2008, stemming from the Lehman Brothers' collapse due to subprime bubbles and their ripple effects, failed 465 U.S. banks. Policymakers introduced a stress test as a standard banking operation to prevent future global fiscal crises. Nowadays, the ongoing bank turmoil shows a different behavior. Researchers found evidence that news and social media might have played a role in catalyzing a bank run by influencing the level of panic (Gunn and Johri 2018; Cookson *et al.* 2023). However, they did not consider the impact of economy-wide variables such as economic policy uncertainty (EPU) and the Federal Reserve Bank's (FRB) monetary policy. Unlike the global fiscal crisis, Fed authorities took immediate control of the failed banks and issued public announcements to alleviate the panic that had spread among their customers (Sandri *et al.* 2023). Economic theory partially explains this pattern, which reflects shifts in economic behavior influenced by evolving popular narratives and interpretations shared among individuals (Shiller 2017).

Although this short-lived bank turmoil caught policymakers by surprise, the root cause may be the savings and loan (S&L) insurance program's outcomes. The existence of the S&L program has undergone severe scrutiny over the past three decades. Kane (1992) argued that regulators allowed insolvent institutions to remain in operation, accumulating additional debt and engaging in speculative strategies that ultimately magnified systemic risk. Such moral hazard eroded profit margins at healthier banks and increased systemic vulnerabilities. At the same time, it remains unclear how the FRB's post-pandemic monetary policy actions, aimed at curbing inflation, have influenced market volatility and EPU. This dynamic interplay among banks, monetary policy, market sentiment, and EPU motivates the current study, which addresses the following research questions:

RQ1. Can dynamic connectedness analysis effectively assess systemic contagion through return and volatility spillovers?
RQ2. What is the impact of market sentiment on the recent U.S. bank turmoil?
RQ3. How does the FRB's monetary policy impact the recent U.S. bank turmoil?
RQ4. What is the impact of EPU on the recent U.S. bank turmoil?

This study makes several contributions to the ensuing discussion. First, it integrates data from multiple sources, including bank stock returns, FRB monetary policy, EPU, and market sentiment, to provide a holistic view of systemic risk. Second, it applies a novel 3C framework—correlation, cointegration, and connectedness analyses—to capture macroeconomic behavior, thereby advancing the academic understanding of financial market dynamics. Third, it employs time-varying analysis on large-scale macroeconomic time series, offering a dynamic and systematic perspective on recent banking sector turmoil.

The study also differs from existing literature in two fundamental ways. Unlike prior work that often examines the economy-wide variables in isolation, this analysis emphasizes their strong interdependence and evaluates their collective influence on bank performance. Furthermore, it treats bank stock returns and economy-wide variables symmetrically by estimating an unconditional time-series process. This approach provides insights into how economy-wide variables react to changes in bank stock performance and, in turn, influence future returns.

The remainder of the paper is organized as follows. Section 2 provides a comprehensive review of risk contagion, market sentiment, spillover, and connectedness of the banking crisis. Section 3 proposes the 3C framework and data collection. Section 4 presents empirical results, while Section 5 discusses the findings and their managerial implications. Finally, Section 6 offers concluding remarks.

## 2. Literature review

Amplified by a highly integrated global financial system, systemic risk has become the focus of many researchers (Mendoza and Quadrini 2010; Brownlees *et al.* 2021; Mohammed *et al.* 2023). There is also a growing trend among them in developing network models for studying financial contagion.

### *2.1 Market sentiment*

While there exists an extensive body of literature discussing the influence of market sentiment on financial market performance, particularly in terms of stock returns (Brown and Cliff 2004; Baker and Wurgler 2006; Tetlock 2007; Carretta *et al.* 2011; Lux 2011; Smales 2017; Gaies *et al.* 2022; Lee *et al.* 2022; Bai *et al.* 2023; Yang 2023; Wang 2025), relatively few studies have focused on its impact on bank performance. Günsür and Bulut (2022) investigated the changing behavior of investors' perceptions of the financial situation of 27 major American bank holding companies during significant events before the 2007 mortgage crisis. Agoraki *et al.* (2022) demonstrated that bank lending decreases in parallel with low investor sentiment, particularly when banks have higher levels of credit risk. Prediction models used by de Jesus and Besarria (2023) exhibit improved accuracy by including a bank sentiment variable when forecasting the risk rating of the banks under scrutiny. Cubillas *et al.* (2021) showed that periods characterized by high investor sentiment positively impact bank lending and foster risk-taking behaviors, potentially compromising bank stability. Anastasiou and Drakos (2021) measured the fear of depositors by proposing search engine-based crisis sentiment indicators and found it to be a significant driver of bank deposit flows across EU countries.

Various methodologies have been developed to measure market sentiment, also called investors' sentiment or attention, reflecting investors' attitudes toward expected returns of specific securities or the entire market (Baker and Wurgler 2007). In 2006, they used a composite proxy for investors' sentiment by constructing the first principal component of six individual sentiment proxies from the U.S. and observed that investors' sentiment has a more substantial impact on small stocks, high-volatility stocks, stocks with low institutional ownership, and stocks that are difficult to arbitrage. Text-based sentiment indices derived from social media and news sources have also gained traction in investors' sentiment analysis (Tetlock 2007; Carretta *et al.* 2011; Garcia 2013; Niţoi and Pochea 2020; Anastasiou and Drakos 2021; Agoraki *et al.* 2022; Bai *et al.*

2023). For instance, Garcia (2013) utilized advanced Natural Language Processing (NLP) algorithms to assess the language and sentiment expressed in social media posts, news articles, and other textual data related to financial markets and assigns sentiment scores to individual pieces of text, indicating the degree of positivity or negativity expressed. Bai *et al.* (2023) compiled a list of financial market-related keywords and analyzed global news reports using the Linguistic Inquiry and Word Count (LIWC) engine to gauge sentiment tendencies. Using an unconventional approach, Edmans *et al.* (2022) explored the relationship between music sentiment derived from music lyrics and stock returns. Additionally, the volatility index (VIX) serves as a widely used sentiment proxy in the U.S. market, facilitating investigations into the relationship between market performance and investors' sentiment (Ben-Rephael *et al.* 2012; Padungsaksawasdi and Daigler 2014; Frijns *et al.* 2016; Smales 2017; Mbanga *et al.* 2019; Günsür and Bulut 2022; Hoekstra and Güler 2024). This study also adopts the VIX as a proxy for market/investor sentiment.

## 2.2 Federal Reserve Bank monetary policy

The FRB's monetary policy is a significant determinant of financial performance (Thorbecke 1997; Rigobon and Sack 2004; Bernanke and Kuttner 2005; Barbosa *et al.* 2018). Central banks use conventional and unconventional monetary policy tools to shape economic conditions and achieve policy objectives (Reis 2013). These tools encompass reserve requirements, interest rate adjustments, open market operations, unconventional measures such as negative interest rates, quantitative easing, forward guidance, and asset purchase programs. For instance, monetary policy changes, such as interest rate hikes or cuts, have significant implications for banks' profitability, liquidity, and lending activities (Stein 2012). Numerous studies have examined how conventional and/or unconventional monetary policies affect banks' performance.

Mamatzakis and Bermpei (2016) uncovered a negative relationship between unconventional monetary policy and bank performance in the U.S. This negative association is moderated for banks with higher degrees of asset diversification and low levels of deposit funding. Borio et al. (2017) investigated the impact of changes in monetary policy stance, particularly interest rate changes, on bank profitability using a sample of international banks. The analysis reveals that monetary policy actions significantly affect bank profitability, with interest rate changes playing a significant role. Chang and Li (2018) examined liquidity effects and monetary policy in a flexible-price model with banking. They found that unexpected money injections impact output and interest rates based on their liquidity relative to existing money stocks. Altavilla *et al.* (2018) examined the relationship between monetary policy and bank profitability in a low-interest-rate environment. They find that accommodative monetary policy, characterized by low interest rates, negatively affects banks' profitability. Demirgüç-Kunt *et al.* (2021) investigated the effect of financial sector policy announcements on bank stocks during the global COVID-19 pandemic. They suggested that liquidity support, monetary easing, and borrower support programs mitigated the negative impact of the bank crisis. Buch *et al.* (2022) looked at the impact of increased competition in banking on the transmission of monetary policy and its impact on economic activity, noting an intensified transmission following banking deregulation. Schonhardt-Bailey *et al.* (2022) employed quantitative text analysis on monetary policy legislative oversight hearings transcripts during the fiscal crisis to compare the UK and U.S. responses. In this study, the Secured Overnight Financing Rate (SOFR), a form of interest rate, is employed as a proxy for the FRB's monetary policy.

## 2.3 Economic policy uncertainty

EPU pertains to the uncertainty of economic policy decisions and their potential economic impacts. This uncertainty stems from various factors, such as political instability, regulatory alterations, and fluctuations in economic conditions (Julio and Yook 2012; Nguyen *et al.* 2024). The increase in economic policy uncertainty following the global financial crisis has underscored the significance of analyzing the impacts of such uncertainties on financial market players, particularly banks (Xu 2020).

However, there is limited research on the effects of EPU on financial institutions' performance, especially in the U.S. financial market. Nguyen *et al.* (2024) investigated the impact of policy-related economic uncertainty on the performance of U.S. bank-holding companies using a comprehensive panel dataset spanning from 2001 to 2021. The empirical results indicate that while EPU negatively affects banking performance, this effect diminishes with increasing levels of bank return on assets (ROA). In contrast, findings from non-U.S. markets are inconsistent. Ozili and Arun (2023) examined the impact of EPU on bank profitability across 22 developed countries. They found that EPU positively influences banks' profitability in Asia and the Americas regions, where a high return on equity was observed during periods of high EPU. In emerging economies, Hamdi and Hassen (2022) found that EPU had a negative effect on the profits of Tunisian commercial banks. Nguyen et al. (2021) observed a positive association between EPU and banks' profitability and cost efficiency.

Studies focusing on the U.S. market primarily use the EPU index developed by Baker *et al.* 2016, which is based on over 2,000 U.S. newspapers and contains a number of uncertainty indicators (Brogaard and Detzel 2015; Baker *et al.* 2016; Bekiros *et al.* 2016; Xu 2020; Tran *et al.* 2021; Nguyen *et al.* 2024). This index is adopted in this study.

## 2.4 Spillover and connectedness analysis for the banking crisis

The TVP-VAR approach employed here allows for analyzing how spillover volatility changes over time and facilitates differentiation between bank stock returns and all three economy-wide variables. Several studies have explored bank performance using VAR models. Using a Granger causality test in a modified VAR model, Ludwig and Sobański (2014) analyzed the direction of bilateral banking industry fragility linkages among Eurozone member countries during the global financial crisis. In contrast, Uch *et al.* (2021) investigated the effect of the banking crisis on credit growth and reported that the adverse impact of banking crises on credit growth is magnified with high levels of financial development. Tabak *et al.* (2022) applied VAR to explore the relationship between banking markets in emerging economies. The results show that the global COVID-19 pandemic influenced the forms and intensity of financial sector connections among countries. Alfarano and Blanco-Arroyo (2022) used VAR to study the impact of banking industry concentration on credit shocks and aggregate credit, and they report that higher banking concentration leads to an increase in the share of bank-specific shocks to aggregate volatility. Mohammed *et al.* (2023) analyzed the interdependence of investors' sentiments (NSS), U.S. interest rates, EPU, and financial assets using quantile VAR-based connectivity and wavelet multiple local correlations. They found that investor sentiments transmit shocks under various market conditions, with fluctuating coefficients over time. While VAR models excel in assessing static connectedness, the latest advancements in TVP-VAR dynamically evaluate the interconnectedness of risks within the financial system.

Several studies investigate the relationships between banks' performance and one of the three economic variables. This is the first to provide a comprehensive framework for investigating the dynamic connectedness between bank performance and market sentiment, FRB's monetary policy, and EPU for a systematic level risk analysis based on the recent U.S. bank turmoil.

## 3. Data and methodology

### *3.1 Data and integration tests*

We selected four failed American banks with the symbols SI, SBNY, SIVB, and FRC, and their four direct competitors with the symbols CMA, ZION, PACW, and WAL, from Yahoo Finance. These competitors were selected based on market capitalization, customer base, and capital structure as reflected in their balance sheets and SEC 10 K reports.

The price information of the bank stocks was collected from the Center for Research in Security Prices and Yahoo Finance. VIX is chosen as the proxy of market sentiment, and VIX's daily information is obtained from the Chicago Board Options Exchange (CBOE). The EPU index is provided by the FRB[2] and is calculated based on daily news from major U.S. newspapers. The monetary policy variable is SOFR, also provided by the FRB[3]. Descriptions of the variables are given in Table 1. March 18, 2022, was the first trading day following the FRB's initial interest rate hike of that cycle, which occurred on March 17, 2022, while March 15, 2023, marked the final trading day for which stock prices of four failed American banks were available. This date is significant as it precedes their collapse.

Researchers utilized the monthly Baker–Bloom–Davis (BBD) index as a proxy for the monetary policy variable (Rossi and Sekhposyan 2015; Iqbal *et al.* 2020). However, our study employs the TVP-VAR model with high-frequency daily data for dynamic connectedness analysis. Thus, the BBD index is deemed unsuitable for this analysis.

**Table 1.** Descriptions of the variables in this study.

| Variable | Symbol | Description | Source |
|---|---|---|---|
| Bank stock returns | SIVB, SI, FRC, CMA, SBNY, PACW, WAL, ZION | Cumulative returns based on the daily adjusted closing price of each bank stock | CRSP and Yahoo Finance |
| Market sentiment | VIX | Volatility index to measure volatility in the stock market accompanied by market fear | CBOE |
| FRB monetary policy | SOFR | Based on overnight, Treasury-backed repo transactions | FRB at New York |
| Economic policy uncertainty | EPU | EPU index based on daily news from major newspapers in the U.S. | FRB at St. Louis |

---

[2] https://fred.stlouisfed.org/series/USEPUINDXD
[3] https://www.newyorkfed.org/markets/reference-rates/additional-information-about-reference-rates

There are five measures of interest rates: Broad General Collateral Rate (BGCR), Effective Federal Funds Rate (EFFR), Overnight Bank Funding Rate (OBFR), Secured Overnight Financing Rate (SOFR), and Tri-party General Collateral Rate (TGCR). The first two are unsecured rates using data on overnight federal funds transactions, while the last three are secured rates using data on overnight repurchase agreements (repo) backed by Treasury Securities. SOFR is the broadest of the secured rates among these five rates[4]. Thus, it is utilized as a proxy for FRB monetary policy.

Table A1 summarizes the bank tickers, assets, liabilities, and loan-deposit-ratio (LDR) in percentage and market caps. The first-order difference of each variable is calculated as the bias correction from seasonal effects or extreme events. These variables are the first-order difference of the bank stocks' cumulative return and the economy-wide variables—VIX, SOFR, and EPU.

Table A2 presents the descriptive statistics of the first-order difference for bank stock returns, the economy-wide variables—VIX, SOFR, and EPU. The means and medians of all variables for the nine banks are negative and close to zero, with UMBF having the highest median value and FRC having the lowest. FRC exhibits more volatility than other banks, as indicated by its significant standard deviation. Negative skewness values suggest a longer left tail in the probability density function, while VIX and SOFR have positive skewness values, indicating longer right tails. EPU has skewness values close to zero, resembling a standard normal distribution. However, the high kurtosis values indicate that these variables have fat tails in their distribution. The Jarque-Bera (JB) normality test rejects normality for the variables at the 1% significance level, other than EPU (Jarque and Bera 1980). The results of the Ljung-Box Q test indicate that all variables have a serial correlation in squared series, suggesting that each variable has a time-varying variance (Ljung and Box 1978). Thus, the TVP-VAR model is appropriate for measuring the connectedness network among the variables.

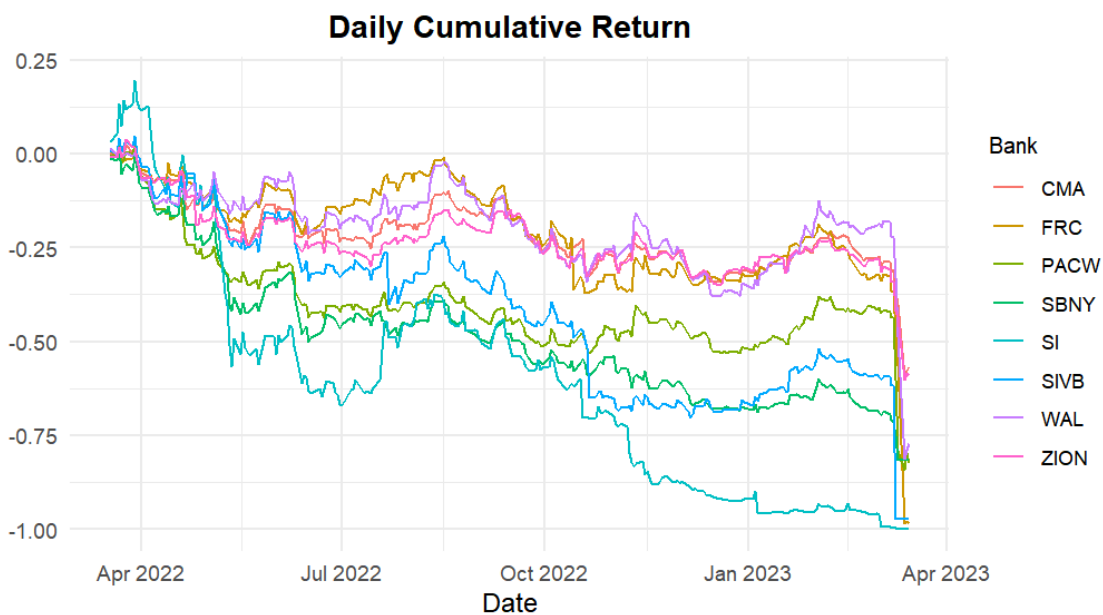

**Figure 1.** Daily cumulative return of bank stocks. Note: The daily cumulative returns of nine banks' portfolios based on the adjusted closing price from March 18, 2022, to March 15, 2023, with a sharp drop on March 9, 2023.

---

[4] https://www.financialresearch.gov/short-term-funding-monitor/

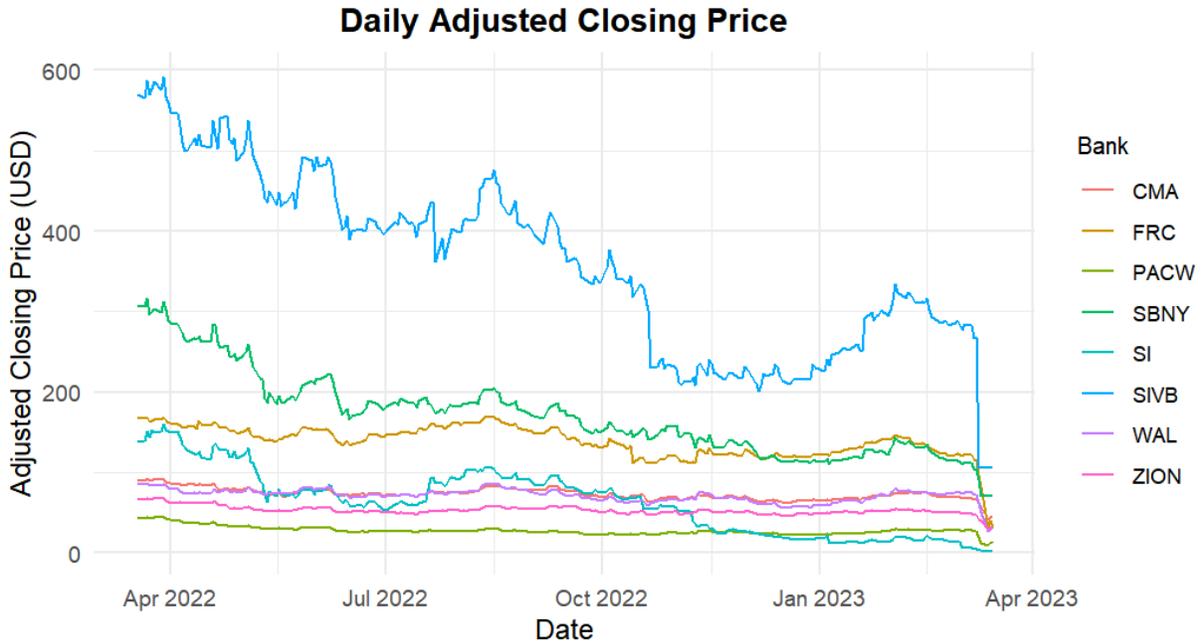

**Figure 2.** Daily adjusted closing price of eight bank stocks from March 18, 2022, to March 15, 2023. Note: A sharp drop on March 9, 2023.

Figures 1 and 2 illustrate the recent performance of the bank stocks in terms of daily cumulative returns and adjusted closing prices from March 18, 2022, to March 15, 2023. The ticker ID is used to label each bank's name. Figure 1 illustrates variations in time-varying volatility across the banks for the given period. Similar fluctuation patterns are observed for all nine banks throughout the period. For example, bank stocks' returns and adjusted closing prices dropped in March 2023. In addition to the descriptive statistics, a test of integration properties of the variables was performed by an augmented Dickey–Fuller (ADF) unit-root test before further analysis, and the results are given in Appendix A.

*3.2 Analytics method*

The 3C framework presented in Figure 3 shows three stages of analytical methods for time-series data: correlation, cointegration, and connectedness. The 3C analyses are implemented in the R language. In the first stage, we report the results of conditional/partial correlation.

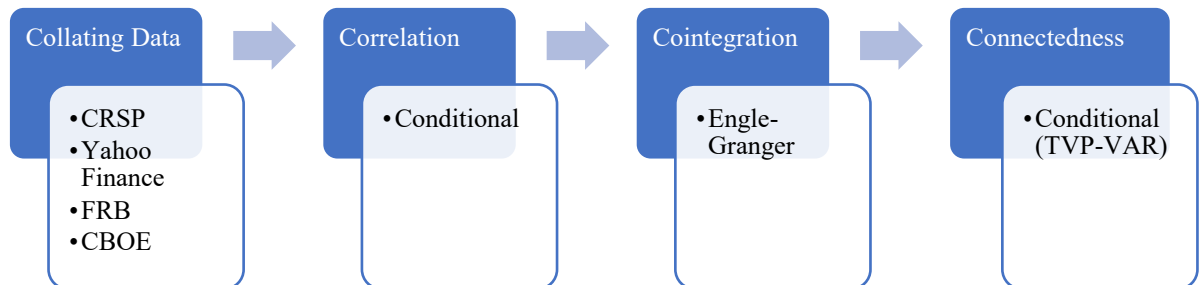

**Figure 3**. Workflow of the 3C framework.

*3.2.1 Correlation*

Conventional static correlation analysis measures the strength and direction of relationships between variables. While the Pearson and Spearman correlations are based on a normal distribution, the Kendall correlation measures the dependency strength. The results of the various correlation measures are offered in the supplemental materials. The conventional static correlation might mislead researchers because it excludes the time-varying factor and assumes a linear relationship between variables. Such correlation changes over time depending on the length of the chosen rolling windows. Therefore, using time-varying estimation, conditional and partial correlation capture changes in correlation over time. Lawrance (1976) explained the difference between conditional and partial correlation and the multivariate distribution where both correlations are equal.

*3.2.2 Cointegration*

The pairwise cointegration analysis in the second stage of 3C provides information about banks' cumulative return spillovers and the connectedness relationship between bank stock returns and the economy-wide variables—market sentiment, FRB monetary policy, and EPU. Figure 4 illustrates the structure of the cointegration analysis among the variables.

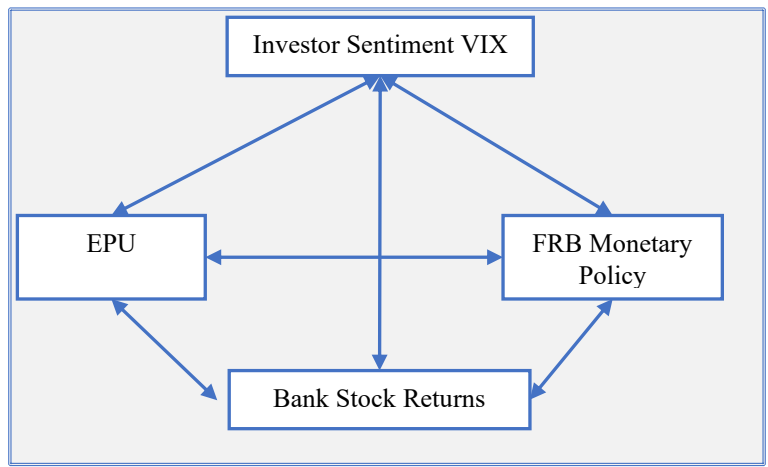

**Figure 4.** Pairwise cointegration analysis.

There are six sets of hypotheses.

H1. There is a strong bidirectional cointegration between market sentiment and EPU.
H2. There is a strong bidirectional cointegration between market sentiment and bank stock returns.
H3. There is a strong bidirectional cointegration between market sentiment and monetary policy.
H4. There is a strong bidirectional cointegration between EPU and monetary policy.
H5. There is a strong bidirectional cointegration between EPU and bank stock returns.
H6. There is a strong bidirectional cointegration between monetary policy and bank stock returns.

The pairwise cointegration analysis is conducted by the augmented Engle-Granger two-step cointegration test (MacKinnon 1994). Its results are validated through the JB goodness-of-fit test, evaluating the distribution of residuals, and are presented in Tables A2 and A6. The *p*-values <0.05 indicate the validity of the cointegration analysis for these variables. A pairwise cointegration between two variables shows the existence of system-wide cointegration.

$$\begin{bmatrix} scr_t \\ inv_t \\ epu_t \\ mpi_t \end{bmatrix} = d_t + \emptyset_1 \begin{bmatrix} scr_{t-1} \\ inv_{t-1} \\ epu_{t-1} \\ mpi_{t-1} \end{bmatrix} + \emptyset_2 \begin{bmatrix} scr_{t-2} \\ inv_{t-2} \\ epu_{t-2} \\ mpi_{t-2} \end{bmatrix} + \cdots + \emptyset_p \begin{bmatrix} scr_{t-p} \\ inv_{t-p} \\ epu_{t-p} \\ mpi_{t-p} \end{bmatrix} + w_t \quad (1)$$

where:
- $scr_t$: Stock cumulative return
- $inv_t$: Market/Investor sentiment index (VIX)
- $epu_t$: Economic policy uncertainty index (EPU)
- $mpi_t$: FRB monetary policy measure (SOFR)

### 3.2.3 Static and dynamic connectedness analysis

The connectedness analysis for the time-series is performed using static and dynamic models. The formulations of both models are presented in Appendix B. A TVP-VAR estimation model was employed to investigate the dynamic conditional connectedness among bank stocks and the economy-wide variables—market sentiment, FRB monetary policy, and EPU. The results of the pairwise connectedness measures, including NPDC, PCI, and PII, reveal the risk contagion within the financial network. The plots of PCI and PII during the period of interest provide insights into the dynamic interconnectedness relationship among the variables as evidence of risk contagion among bank returns.

## 4. Empirical results

### 4.1 Conditional and partial correlation of VAR between stock returns and the economy-wide variables

Tables A4 and A5 in the Appendix report the conditional correlation and partial correlation using the VAR model between the stock returns of the nine banks and the economy-wide variables. The conditional correlation in Table A4 shows that market sentiment presents a negative relationship with stock returns but a positive relationship with monetary policy. The FRB's monetary policy has a negative relationship with stock returns, while EPU has a positive relationship with stock returns. The relationships between stock returns are all positive.

Table A5 shows the partial correlations, and economy-wide variables have mixed and weak relationships with stock returns. The relationships between a pair of stock returns are also mixed and weak.

The values of the conditional correlation coefficients of each pair of variables are generally higher than their partial correlation coefficients. Therefore, we considered the TVP-VAR when dealing with the dynamic nonlinear relationship of time-series data at first-order difference.

## 4.2 Pairwise cointegration and the Chow test

The augmented Engle-Granger two-step cointegration test was performed, and the pairwise cointegration test results are presented in the Appendix, Table A6. All variables have an extremely strong bidirectional cointegration relationship. Thus, all hypotheses (H1-H6 in Section 3.2.2) are supported by the evidence.

The Chow test for parameter stability reports structural changes in Table 2. The null hypothesis of the Chow test is rejected with a $p$-value of 0.002. Thus, the breakpoint test shows the instabilities of the estimated parameters in the linear model. The Chow test supports using the TVP-VAR model as a better method for the nonlinearity of stochastic volatility.

**Table 2.** Chow test on structural stability of VAR model.

|  | Test statistic | 95% critical value | $p$-value |
|---|---|---|---|
| Break-point test | 3577.2 | 3483 | 0.002 ** |
| Sample-split test | 898.3 | 887 | 0.030 * |

Note: $p$-value<=0.1, * $p$-value<=0.05, ** $p$-value<=0.01, and *** $p$-value<=0.005.

## 4.3 Dynamic conditional connectedness

The Chow test identified structural instabilities in the system, suggesting that the standard linear VAR model might not fully capture the nonlinear and time-varying relationships among variables. While static connectedness measures from the VAR provide a broad overview of spillovers, a dynamic framework offers more nuanced insights into how these relationships evolve over time.

Using the total connectedness index (TCI) as a benchmark, the static connectedness analysis in Table 3 shows that, on average, about 67% of the forecast error variance of each variable can be explained by spillovers from other variables, underscoring the high level of interconnectedness in the network. Among the banks, PACW, WAL, SIVB, SBNY, and ZION are the primary net transmitters (givers) of shocks, with PACW (+22.3%) and WAL (+12.4%) contributing the most. In contrast, SI, VIX, FRC, SOFR, and EPU act as net receivers, with SI (–19.8%) and VIX (–15.3%) absorbing the largest share of spillovers. Bold lines in Figure 5a highlight strong spillovers between the failed banks (SI and SBNY), SOFR, and EPU, confirming the high interconnectedness of these nodes.

However, the static analysis does not reveal how spillovers vary over time. Figure 5b addresses this by showing that spillovers are concentrated in specific episodes rather than being constant. For example, SIVB and FRC display positive NET spikes during stress periods, while SI consistently shows negative NET spikes, and SBNY transitions from strong negative NET spikes before July 2022 to muted positive spikes after October 2022. PACW similarly shifts from a net receiver before July 2022 to a low-level transmitter thereafter. Among broader variables, SOFR exhibits strong positive NET spikes before July 2022 but becomes a marginal receiver afterward, while EPU remains a consistent net receiver and VIX alternates between positive and negative NET spikes. These patterns highlight the episodic nature of risk transmission in the network.

**Table 3.** VAR connectedness with time.

|  | SIVB | SI | SBNY | FRC | WAL | PACW | CMA | ZION | VIX | SOFR | EPU | Receiver |
|---|---|---|---|---|---|---|---|---|---|---|---|---|
| SIVB | 27.54 | 4.98 | 9.99 | 8.71 | 11.15 | 16.05 | 7.38 | 9.36 | 4.76 | 0.06 | 0.02 | 72.46 |
| SI | 6.17 | 34.78 | 16.72 | 4.15 | 6.94 | 7.02 | 5.87 | 5.9 | 11.66 | 0.38 | 0.42 | 65.22 |
| SBNY | 8.47 | 10.83 | 22.64 | 7.42 | 10.68 | 12.7 | 9.19 | 9.77 | 7.81 | 0.22 | 0.27 | 77.36 |
| FRC | 16.12 | 2.41 | 5.95 | 19.76 | 13.92 | 15.03 | 10.08 | 12.23 | 3.81 | 0.14 | 0.55 | 80.24 |
| WAL | 13.33 | 3.6 | 7.95 | 11.77 | 18.56 | 15.85 | 11.28 | 12.77 | 4.27 | 0.08 | 0.55 | 81.44 |
| PACW | 14.48 | 4.14 | 10.45 | 8.68 | 13.35 | 22.12 | 10.08 | 11.8 | 4.37 | 0.16 | 0.37 | 77.88 |
| CMA | 9.41 | 3.52 | 8.06 | 10.18 | 13.87 | 12.26 | 20.37 | 15.96 | 5.37 | 0.24 | 0.76 | 79.63 |
| ZION | 8.49 | 3.56 | 8.32 | 11.06 | 14.05 | 12.59 | 15.37 | 19.67 | 6.09 | 0.14 | 0.67 | 80.33 |
| VIX | 5.62 | 10.79 | 10.56 | 6.46 | 7.65 | 7.02 | 8.38 | 9.56 | 30.92 | 0.66 | 2.37 | 69.08 |
| SOFR | 0.37 | 0.63 | 0.99 | 0.18 | 0.18 | 0.54 | 0.86 | 0.5 | 3.4 | 91.93 | 2.09 | 8.07 |
| EPU | 0.38 | 0.97 | 1.37 | 1.63 | 2.05 | 1.14 | 1.82 | 1.35 | 2.22 | 2.09 | 84.97 | 15.03 |
| Giver | 82.83 | 45.42 | 80.36 | 70.23 | 93.85 | 100.21 | 80.31 | 89.18 | 53.77 | 4.17 | 6.41 |  |
| Inc.Own | 110.37 | 80.2 | 103 | 89.99 | 112.41 | 122.33 | 100.68 | 108.85 | 84.69 | 96.1 | 91.38 | TCI |
| NET | 10.38 | -19.8 | 3 | -10.01 | 12.41 | 22.33 | 0.68 | 8.85 | -15.31 | -3.9 | -8.62 | 64.25 |
| NPT | 7 | 2 | 5 | 6 | 8 | 10 | 5 | 6 | 2 | 1 | 1 |  |

Note: Givers: SIVB, SBNY, WAL, PACW, CMA, and ZION. Receivers: SI, FRC, VIX, SOFR, and EPU. The table reports the pairwise variance decomposition shares, Receivers, Givers, total spillover including own variance (Inc.Own), net transmission (NET), and net pairwise transmissions (NPT) for each bank, market sentiment (VIX), monetary policy (SOFR), and economic policy uncertainty (EPU).

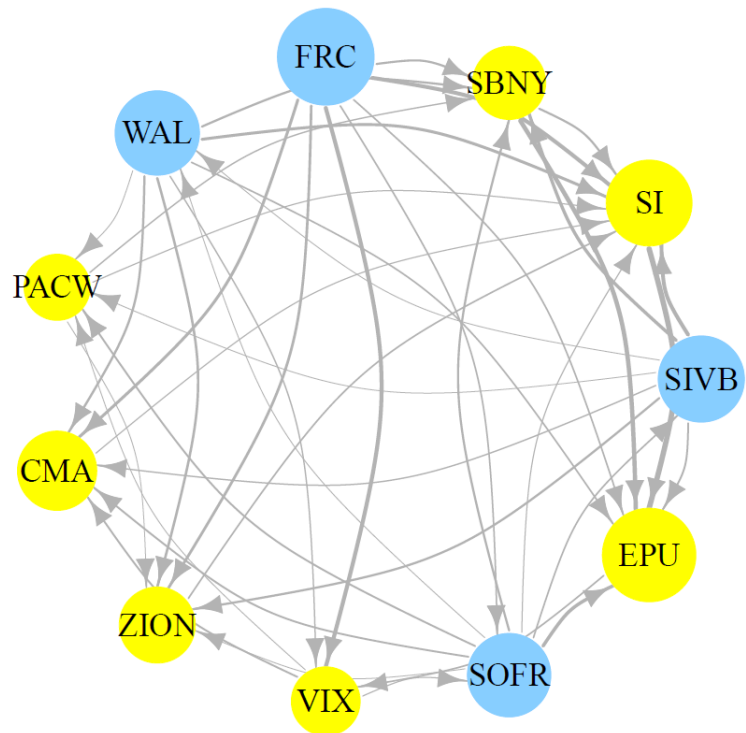

**Figure 5a.** NPDC measure plot of TVP-VAR connectedness network. Note: The graphic results show the level of dynamic conditional connectedness across all pairs in the financial network with the nodes in blue as net giver and the nodes in yellow as net receiver of an uncertainty shock. The arrow in the graph indicates greater directional connectedness in terms of time-varying pairwise net connectedness. The size of the nodes is measured by the weighted time-varying net total directional connectedness. The bold line indicates the level of spillover higher than fine lines between variables over time.

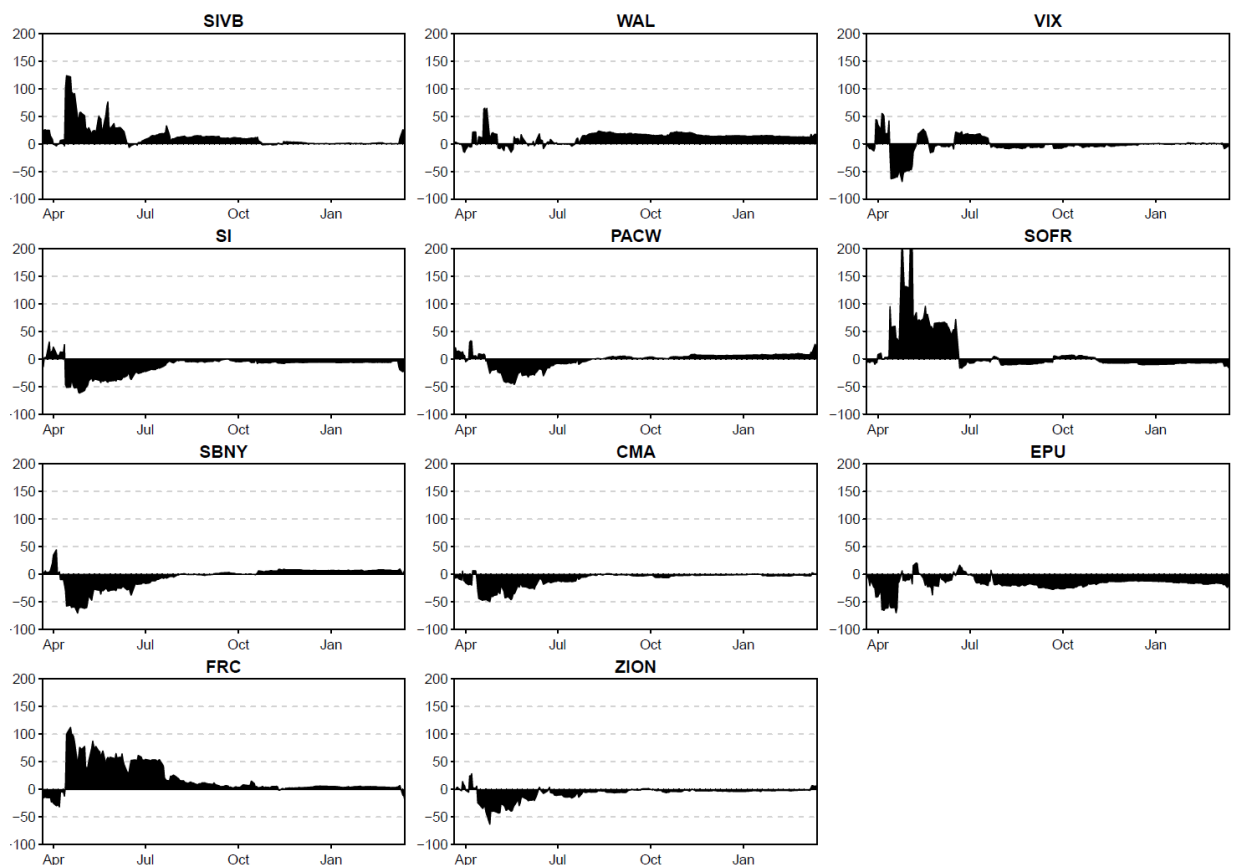

**Figure 5b.** Net volatility spillovers plot of the TVP-VAR model. Note: Shades in the positive value range indicate the giving end of net volatility transmissions and vice versa.

To explore pairwise relationships in greater depth, Figures C1–C11 (Appendix C), D1–D11 (Appendix D), and E1–E11 (Appendix E) illustrate the detailed spillover and influence indices. These plots show that pairwise connectedness is typically higher among bank returns than between banks and SOFR, indicating significant risk contagion within the banking sector. Failed banks exhibit fewer spillovers to SOFR but stronger spillovers to VIX and EPU during the May–July 2022 stress window. Importantly, economy-wide variables exert greater influence on banks than the reverse, underscoring their systemic importance.

Building on this, three TVP-VAR dynamic conditional connectedness analyses were conducted to evaluate the time-varying impact of each macro-financial variable (VIX, SOFR, and EPU) on bank stock returns.

Table 4 and Figure 6a present the results for VIX. The TCI is 77.1%, meaning that over three-quarters of the forecast error variance is attributable to spillovers within the system. SIVB, FRC, and WAL are the major net transmitters (blue nodes) with positive NET values (15.2%, 19.6%, and 12.8%, respectively), while SI, SBNY, PACW, CMA, ZION, and the VIX are net receivers (yellow nodes). SI is the most vulnerable net receiver with a NET value of –18.7%. Figure 6b shows that SIVB and FRC's transmitter roles were concentrated before October 2022, while SI and SBNY remained receivers throughout most of the period, though SBNY transitioned to weak positive NET values after October 2022. PACW and VIX alternate between roles, and CMA and ZION remain consistent net receivers.

**Table 4.** TVP-VAR dynamic connectedness analysis with stock returns and VIX.

|  | SIVB | SI | SBNY | FRC | WAL | PACW | CMA | ZION | VIX | Receiver |
|---|---|---|---|---|---|---|---|---|---|---|
| SIVB | 27.77 | 6.34 | 10.75 | 8.95 | 11.7 | 12.32 | 8.53 | 7.66 | 5.99 | 72.23 |
| SI | 10.73 | 32.36 | 14.25 | 8.29 | 8.54 | 7.57 | 4.98 | 5.41 | 7.87 | 67.64 |
| SBNY | 14.54 | 12.23 | 15.81 | 7.82 | 12.55 | 15.5 | 9.42 | 7 | 5.13 | 84.19 |
| FRC | 8.97 | 3.76 | 5.19 | 34.85 | 11.72 | 7.11 | 6.77 | 9.11 | 12.52 | 65.15 |
| WAL | 12.89 | 5.33 | 10.1 | 11.82 | 14.86 | 13.51 | 12.57 | 12.68 | 6.23 | 85.14 |
| PACW | 13.73 | 6.34 | 13.81 | 7.84 | 14.82 | 13.8 | 14.01 | 10.6 | 5.05 | 86.2 |
| CMA | 10.3 | 3.5 | 9 | 10.12 | 15.09 | 14.39 | 15.18 | 15.4 | 7.02 | 84.82 |
| ZION | 10.12 | 3.43 | 7.11 | 12.2 | 15.4 | 11.48 | 15.28 | 15.33 | 9.65 | 84.67 |
| VIX | 6.12 | 8.03 | 4.65 | 17.73 | 8.08 | 4.22 | 5.09 | 10.2 | 35.87 | 64.13 |
| Giver | 87.4 | 48.97 | 74.87 | 84.77 | 97.9 | 86.09 | 76.67 | 78.06 | 59.45 | 694.18 |
| Inc.Own | 115.16 | 81.32 | 90.67 | 119.62 | 112.76 | 99.89 | 91.85 | 93.39 | 95.32 | TCI |
| NET | 15.16 | –18.68 | –9.33 | 19.62 | 12.76 | –0.11 | –8.15 | –6.61 | –4.68 | 77.13 |
| NPT | 8 | 1 | 2 | 7 | 6 | 4 | 2 | 3 | 3 |  |

Note: Givers: SIVB, FRC, and WAL. Receivers: SI, SBNY, PACW, CMA, ZION, and VIX.

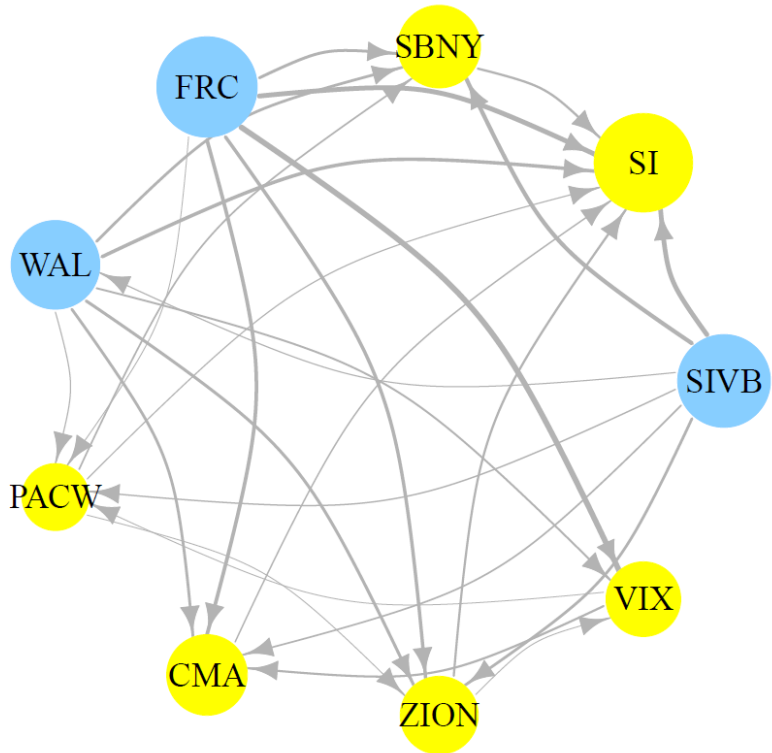

**Figure 6a.** NPDC measure plot of TVP-VAR conditional connectedness network between bank stock returns and VIX.

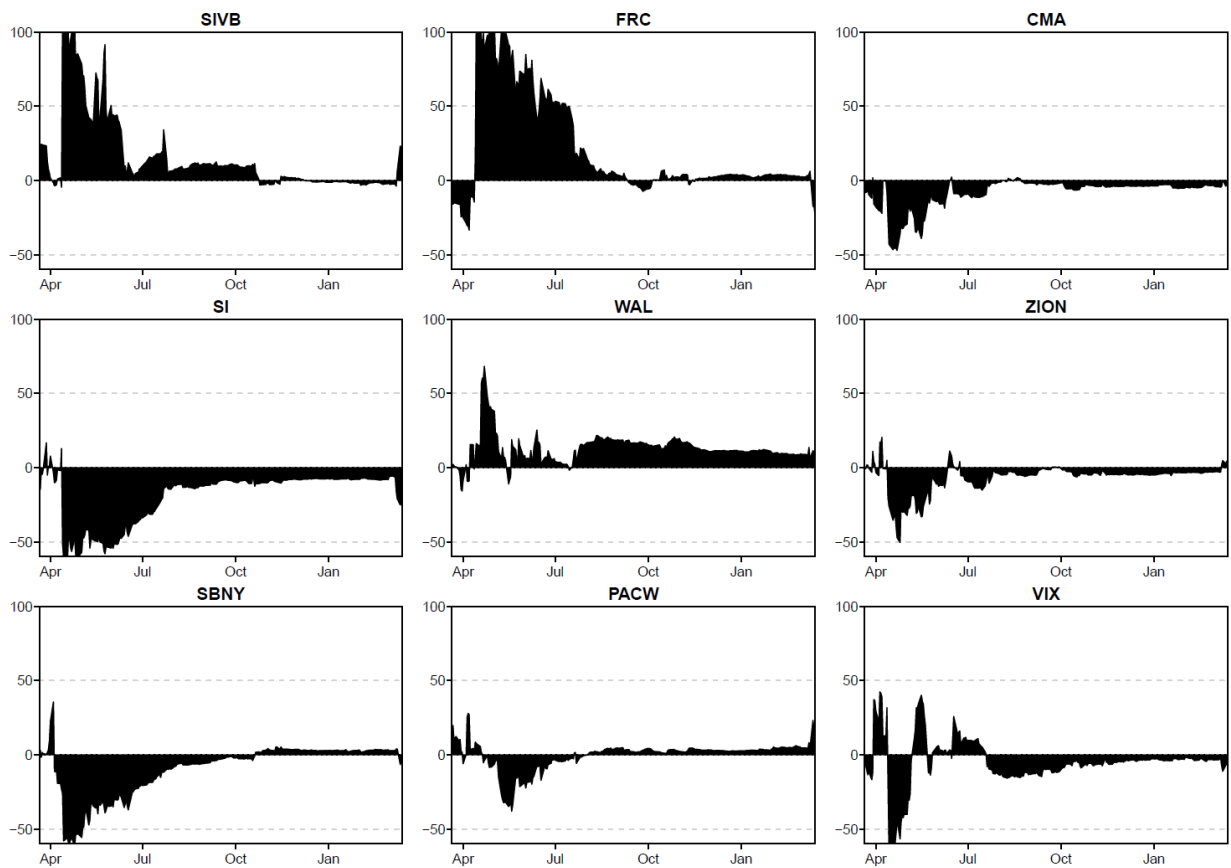

**Figure 6b.** Net volatility spillovers plot of the TVP-VAR model conditional connectedness between bank stock returns and VIX.

Figures 6c–6e build on the network and time-varying connectedness patterns observed in Figures 6a and 6b by focusing specifically on the directional spillovers between the VIX and bank stock returns. The PCI (Figure 6c) quantifies the overall strength of these spillovers over time, the PII (Figure 6d) shows the information share contributed by each variable to the connectedness with the VIX, and the NPDC plot (Figure 6e) identifies which banks act as net transmitters or receivers in their pairwise relationships with the VIX. The patterns observed in Figures 6c–6e further emphasize that the spillovers between the VIX and bank stock returns are episodic and asymmetric. The PCI (Figure 6c) and PII (Figure 6d) plots display predominantly positive spikes with right-skewed shapes, indicating that the intensity and information share of connectedness surge sharply during stress periods rather than remaining constant over time. These spikes reflect infrequent but high-magnitude spillover episodes, consistent with systemic risk dynamics where market sentiment and bank returns become tightly coupled during crises. Similarly, the NPDC plot (Figure 6e) shows mostly positive spikes, suggesting that banks more often act as net transmitters of shocks relative to VIX during these periods. Together, these results show that the network's vulnerability is concentrated in distinct episodes of market stress, driven largely by key banks amplifying shocks outward to the broader system and market sentiment.

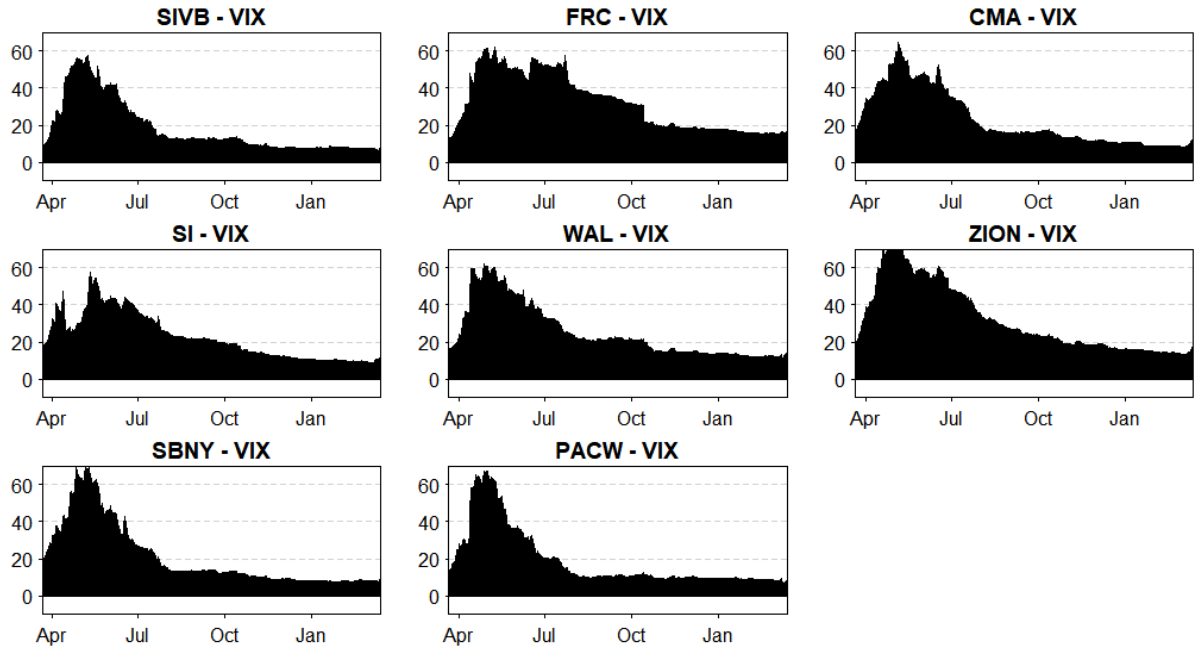

**Figure 6c.** PCI plot between VIX and bank returns. Note: The PCI values between VIX and bank returns from May 2022 to July 2022 are higher than in other periods.

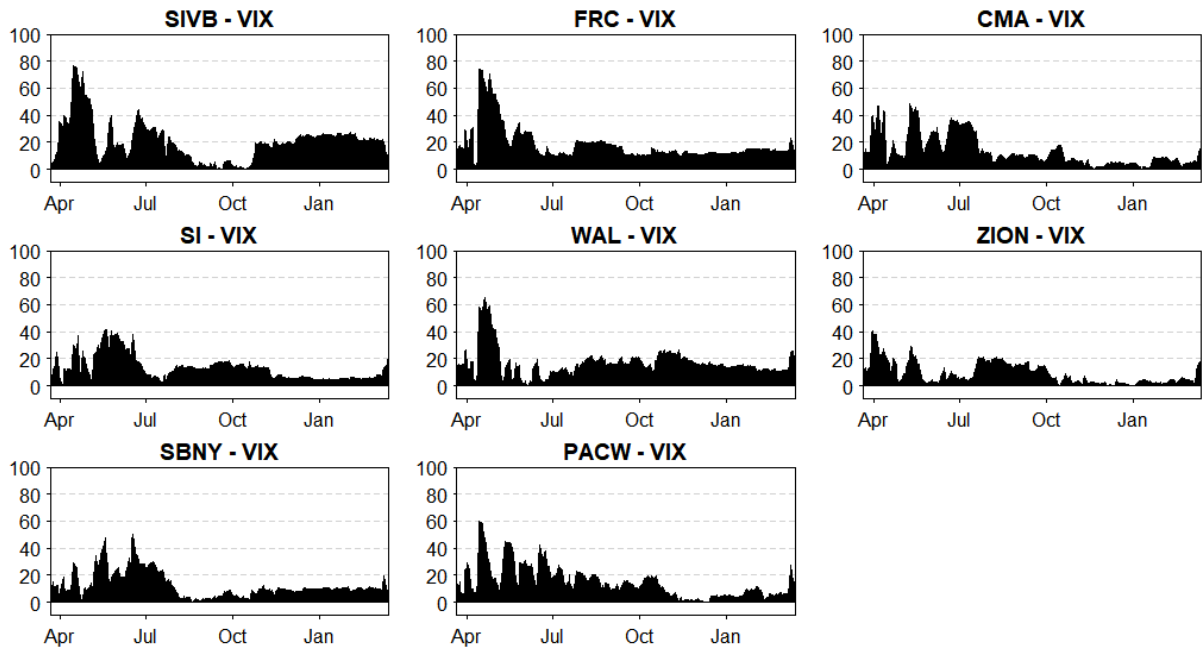

**Figure 6d.** PII plot between VIX and bank returns. Note: The PII values between VIX and bank returns show a sharp increase in March 2023.

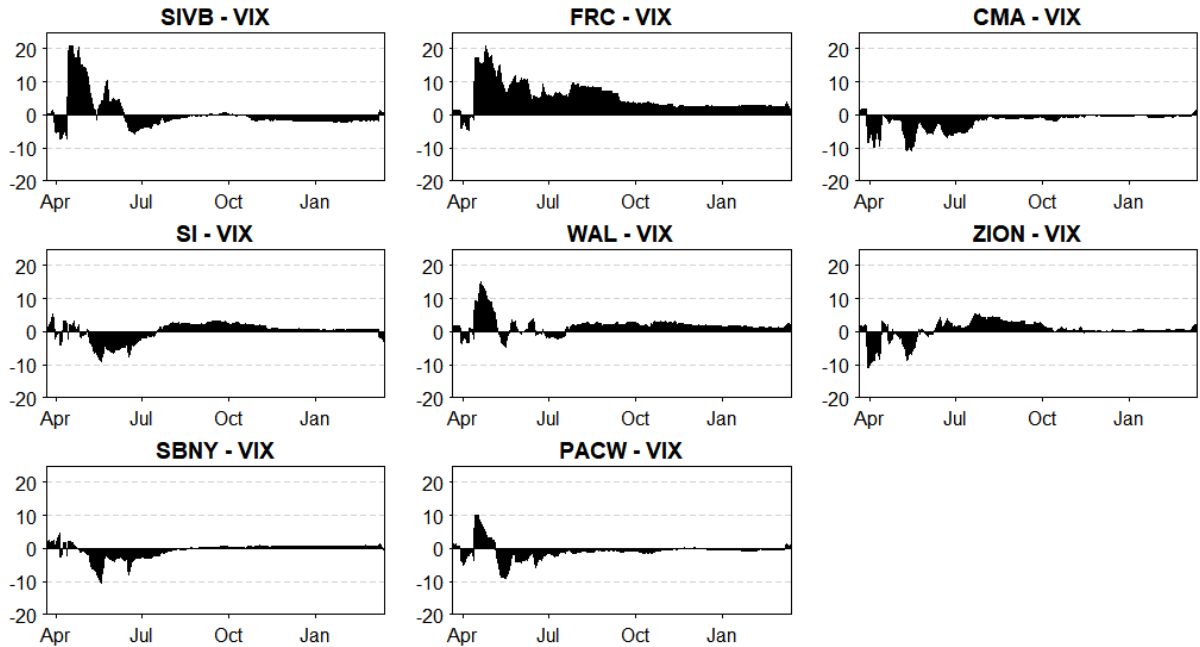

**Figure 6e.** NPDC plot between VIX and bank returns.

Similarly, Table 5 and Figure 7a show dynamic connectedness with SOFR. Here, SOFR, SIVB, FRC, and WAL are the dominant net transmitters (blue nodes), with SOFR having the largest NET value (+11.3%), while SI, SBNY, PACW, CMA, and ZION are net receivers (yellow nodes), with SI showing the largest negative NET (–22.0%). Figure 7b highlights that SOFR and key banks like SIVB and FRC were strongest transmitters prior to July–October 2022 but weakened thereafter, whereas SI and SBNY remained persistent net receivers until SBNY's NET values turned slightly positive late in the period.

**Table 5.** TVP-VAR dynamic connectedness analysis with stock returns and SOFR.

|  | SIVB | SI | SBNY | FRC | WAL | PACW | CMA | ZION | SOFR | Receivers |
|---|---|---|---|---|---|---|---|---|---|---|
| SIVB | 28.73 | 6.6 | 11.16 | 9.06 | 12.03 | 12.74 | 8.84 | 7.91 | 2.94 | 71.27 |
| SI | 11.23 | 34.39 | 15.21 | 8.59 | 8.98 | 8 | 5.27 | 5.73 | 2.6 | 65.61 |
| SBNY | 14.67 | 12.53 | 16.3 | 7.73 | 12.74 | 15.82 | 9.62 | 7.15 | 3.45 | 83.7 |
| FRC | 9.65 | 4.2 | 5.71 | 38.86 | 12.89 | 7.72 | 7.45 | 10.1 | 3.42 | 61.14 |
| WAL | 13.27 | 5.57 | 10.53 | 12.13 | 15.44 | 14.04 | 13.1 | 13.2 | 2.73 | 84.56 |
| PACW | 13.88 | 6.46 | 14.11 | 7.83 | 15.06 | 14.09 | 14.26 | 10.79 | 3.52 | 85.91 |
| CMA | 10.55 | 3.64 | 9.33 | 10.33 | 15.57 | 14.9 | 15.71 | 15.97 | 4 | 84.29 |
| ZION | 10.68 | 3.7 | 7.62 | 12.95 | 16.42 | 12.29 | 16.42 | 16.51 | 3.41 | 83.49 |
| SOFR | 1.04 | 0.95 | 0.98 | 5 | 1.59 | 1.22 | 1.63 | 2.38 | 85.21 | 14.79 |
| Giver | 84.96 | 43.64 | 74.66 | 73.62 | 95.29 | 86.72 | 76.57 | 73.22 | 26.07 | 634.75 |
| Inc.Own | 113.69 | 78.03 | 90.95 | 112.49 | 110.73 | 100.81 | 92.28 | 89.73 | 111.29 | TCI |
| NET | 13.69 | –21.97 | –9.05 | 12.49 | 10.73 | 0.81 | –7.72 | –10.27 | 11.29 | 70.53 |
| NPT | 7 | 0 | 2 | 6 | 6 | 4 | 3 | 1 | 7 |  |

Note: Givers: SIVB, FRC, WAL, PACW, and SOFR. Receivers: SI, SBNY, CMA, and ZION.

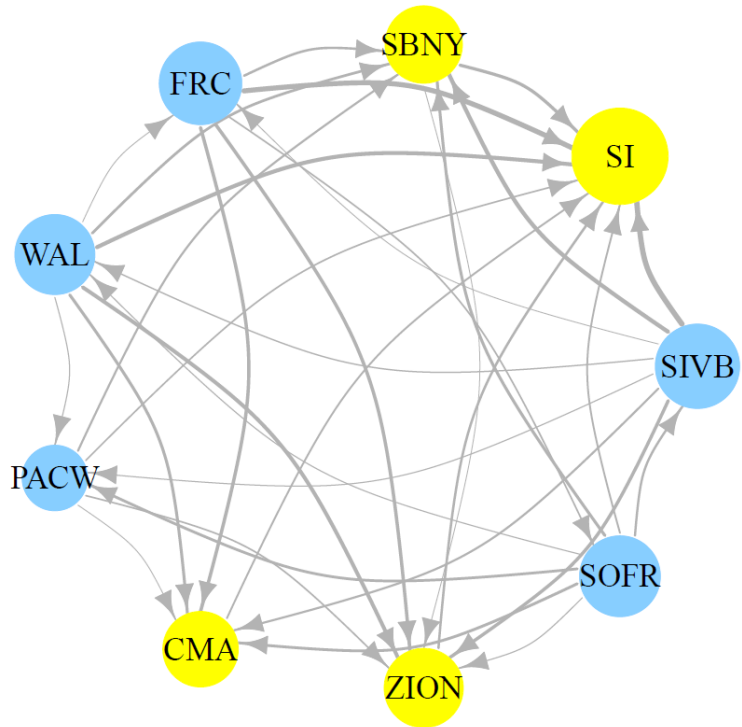

**Figure 7a.** NPDC measure plot of TVP-VAR conditional connectedness network between bank stock returns and monetary policy (SOFR).

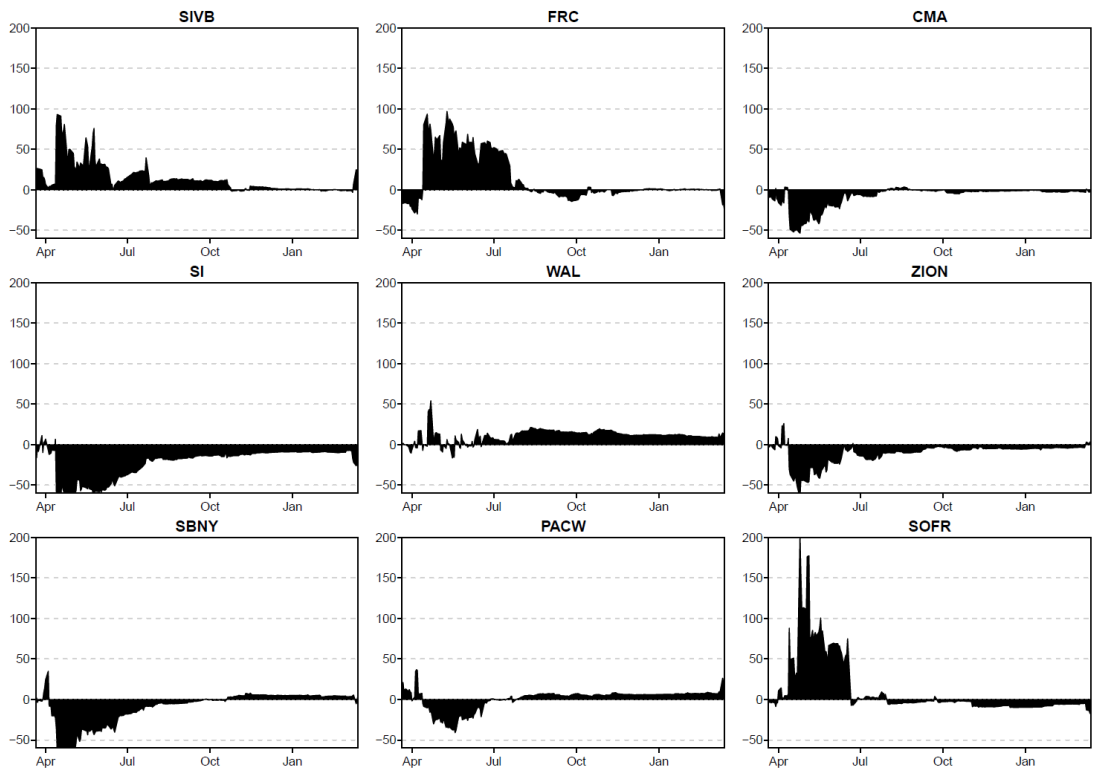

**Figure 7b.** Net volatility spillovers plot of the TVP-VAR model conditional connectedness between bank stock returns and SOFR.

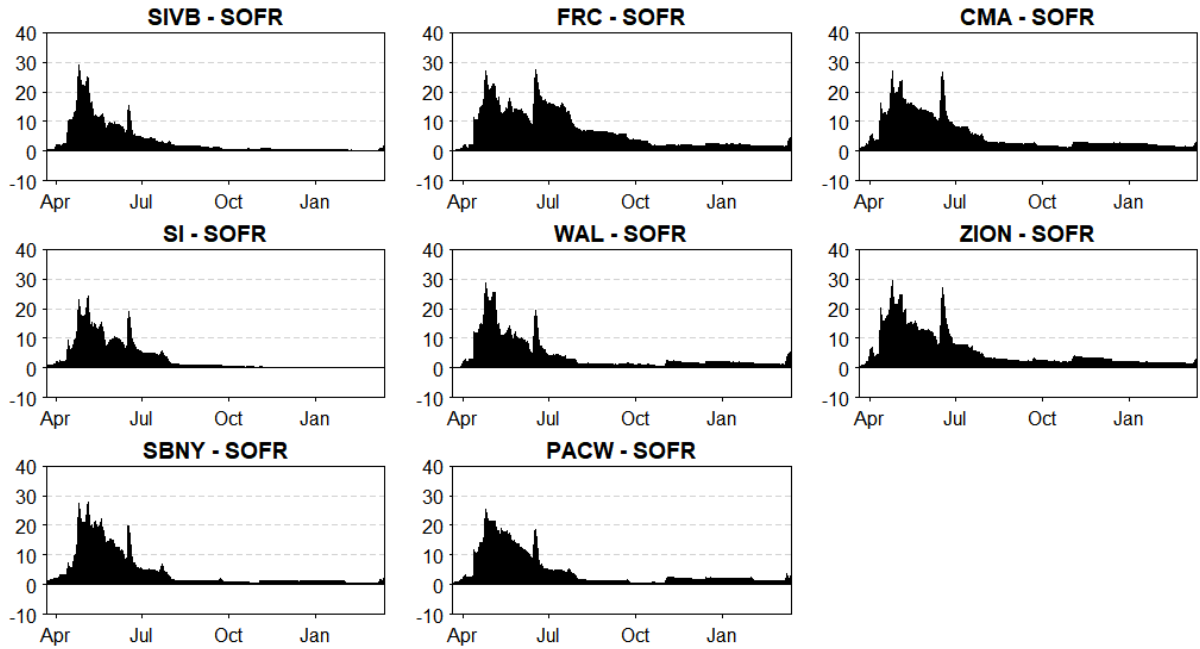

**Figure 7c.** PCI plot between SOFR and bank returns. Note: The PCI values between VIX and bank returns from May 2022 to August 2022 are higher than in other periods.

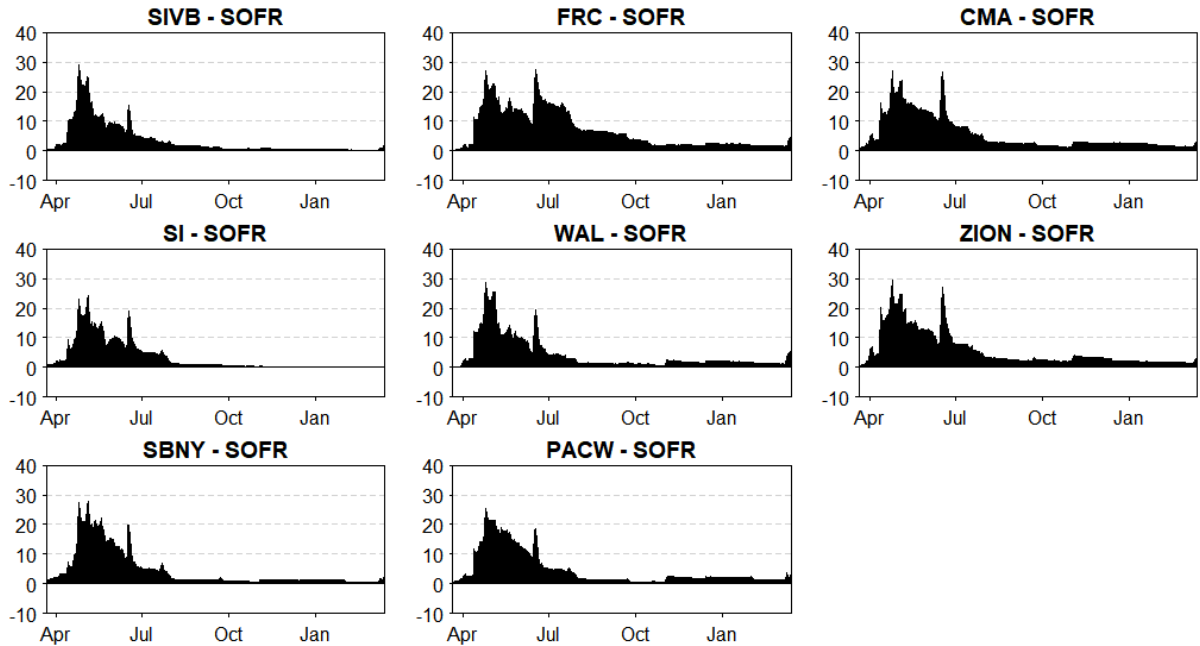

**Figure 7d.** PII plot between SOFR and bank returns. Note: The PII values between SOFR and bank returns have higher values between November 2022 and March 2023.

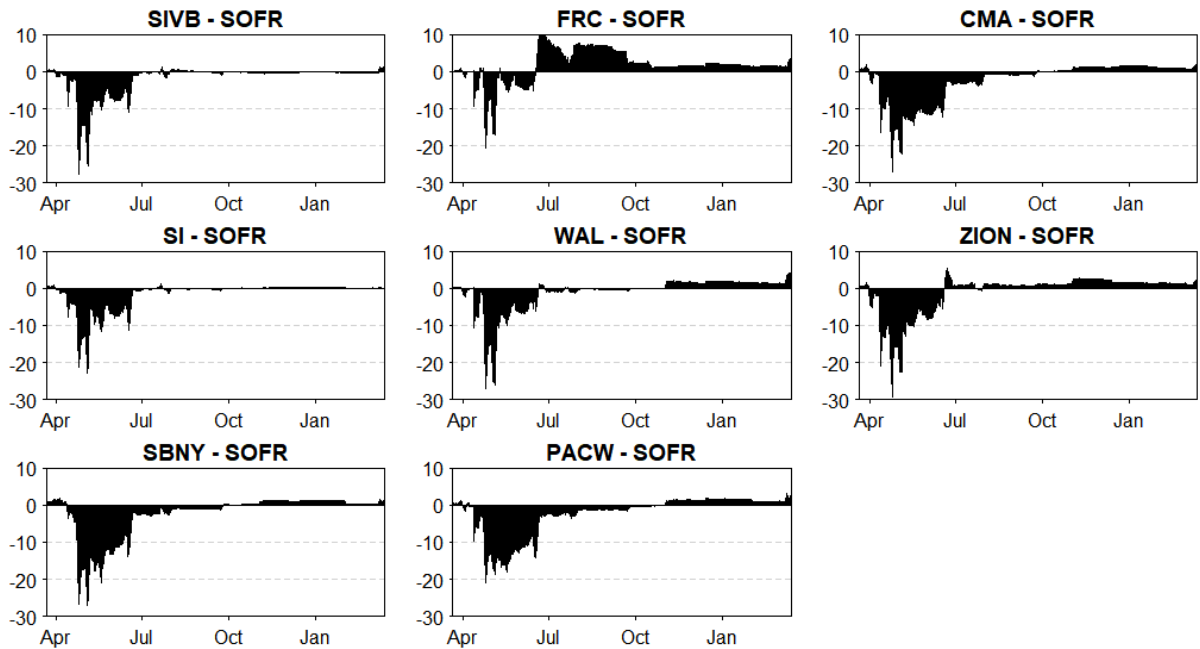

**Figure 7e.** NPDC plot between SOFR and bank returns. Note: The NPDC values between SOFR and failed bank returns have negative values between May and July 2022.

Figures 7c–7e again show predominantly positive spikes, indicating elevated spillover intensity and information sharing during stress periods. The NPDC values in Figure 7e are mostly negative prior to July 2022, suggesting that banks were absorbing shocks relative to SOFR but shift toward low positive levels afterwards, reflecting the diminishing transmitter role of SOFR. These results confirm that SOFR was a key driver of systemic risk propagation in early 2022, but its influence waned in later months.

**Table 6.** TVP-VAR dynamic connectedness analysis with stock returns and EPU.

|  | SIVB | SI | SBNY | FRC | WAL | PACW | CMA | ZION | EPU | Receiver |
|---|---|---|---|---|---|---|---|---|---|---|
| SIVB | 28.88 | 6.61 | 11.19 | 9.23 | 12.13 | 12.8 | 8.87 | 7.96 | 2.33 | 71.12 |
| SI | 11.14 | 34.08 | 15 | 8.52 | 8.92 | 7.93 | 5.2 | 5.66 | 3.54 | 65.92 |
| SBNY | 14.86 | 12.56 | 16.22 | 7.98 | 12.84 | 15.87 | 9.65 | 7.18 | 2.83 | 83.78 |
| FRC | 9.89 | 4.25 | 5.79 | 39.25 | 13.09 | 7.85 | 7.56 | 10.22 | 2.11 | 60.75 |
| WAL | 13.43 | 5.58 | 10.56 | 12.32 | 15.51 | 14.11 | 13.14 | 13.25 | 2.1 | 84.49 |
| PACW | 14.02 | 6.5 | 14.18 | 7.97 | 15.17 | 14.16 | 14.35 | 10.87 | 2.78 | 85.84 |
| CMA | 10.84 | 3.7 | 9.48 | 10.69 | 15.89 | 15.16 | 15.96 | 16.25 | 2.02 | 84.04 |
| ZION | 10.88 | 3.72 | 7.68 | 13.16 | 16.58 | 12.38 | 16.5 | 16.54 | 2.57 | 83.46 |
| EPU | 4.28 | 9.34 | 7.43 | 3.83 | 3.86 | 2.79 | 2.3 | 2.27 | 63.9 | 36.1 |
| Giver | 89.35 | 52.26 | 81.3 | 73.7 | 98.48 | 88.89 | 77.58 | 73.65 | 20.28 | 655.49 |
| Inc.Own | 118.23 | 86.34 | 97.53 | 112.95 | 113.99 | 103.05 | 93.54 | 90.2 | 84.18 | TCI |
| NET | 18.23 | –13.66 | –2.47 | 12.95 | 13.99 | 3.05 | –6.46 | –9.80 | –15.82 | 72.83 |
| NPT | 8 | 1 | 3 | 6 | 7 | 5 | 4 | 1 | 1 |  |

Note: Givers: SIVB, FRC, WAL, and PACW. Receivers: SI, SBNY, CMA, ZION, and EPU.

Finally, Table 6 and Figure 8a present the dynamic connectedness with EPU. SIVB, FRC, WAL, and PACW are net transmitters (blue nodes), while SI, SBNY, CMA, ZION, and EPU are net receivers (yellow nodes), with EPU and SI showing the most negative NET values (–15.8% and –13.7%, respectively). Figure 8b shows that SIVB and FRC's transmitter roles were concentrated prior to October 2022, while SI, SBNY, CMA, and ZION consistently absorbed shocks. EPU itself remained a net receiver throughout the sample.

Figures 8c–8e (PCI, PII, and NPDC) show that connectedness between EPU and bank returns intensified during stress episodes, as reflected by the uniformly positive spikes. These findings demonstrate that EPU primarily absorbs shocks rather than propagating them, but its ties to the banking system strengthen sharply during periods of elevated systemic risk, especially before October 2022.

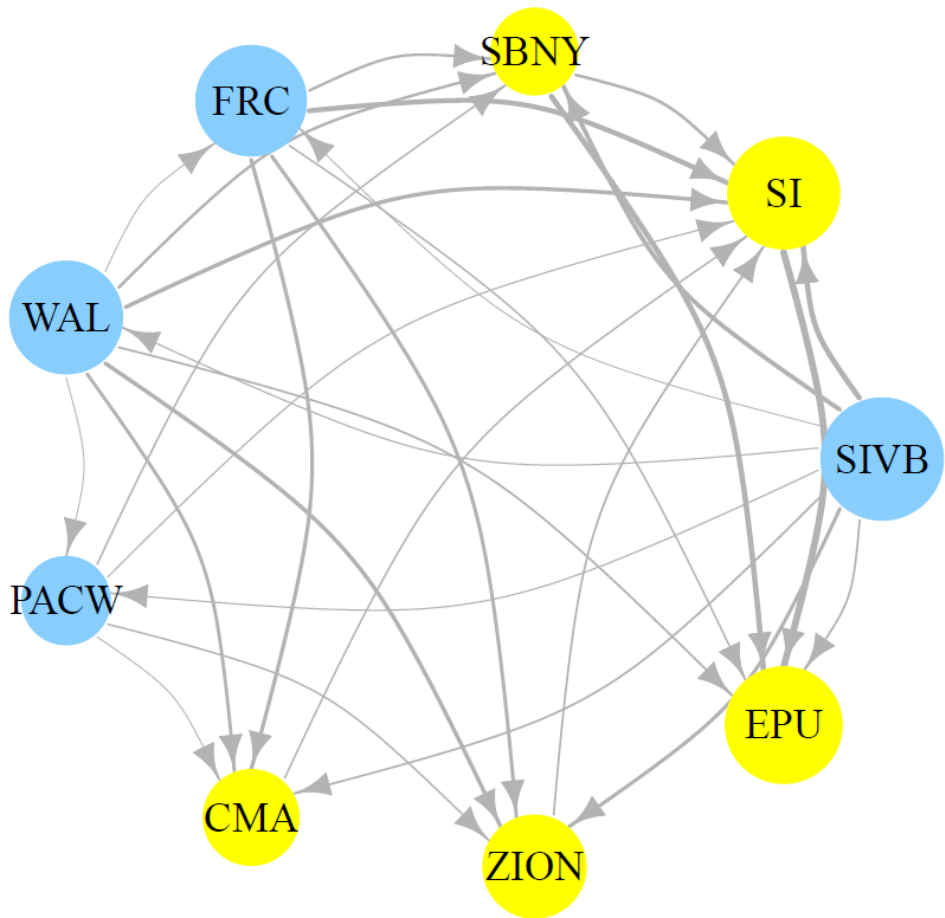

**Figure 8a.** NPDC measure plot of TVP-VAR conditional connectedness network between bank stock returns and EPU.

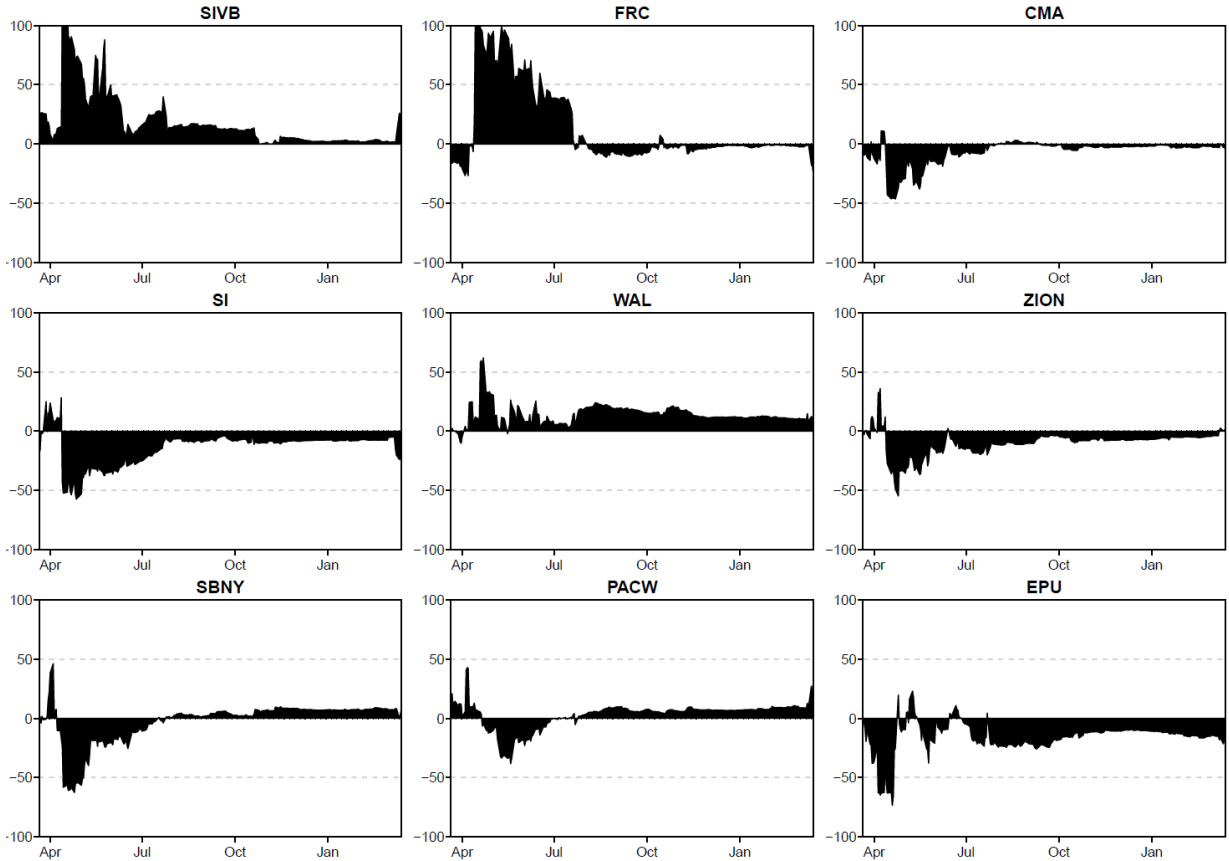

**Figure 8b.** Net volatility spillovers plot of the TVP-VAR model conditional connectedness between bank stock returns and EPU.

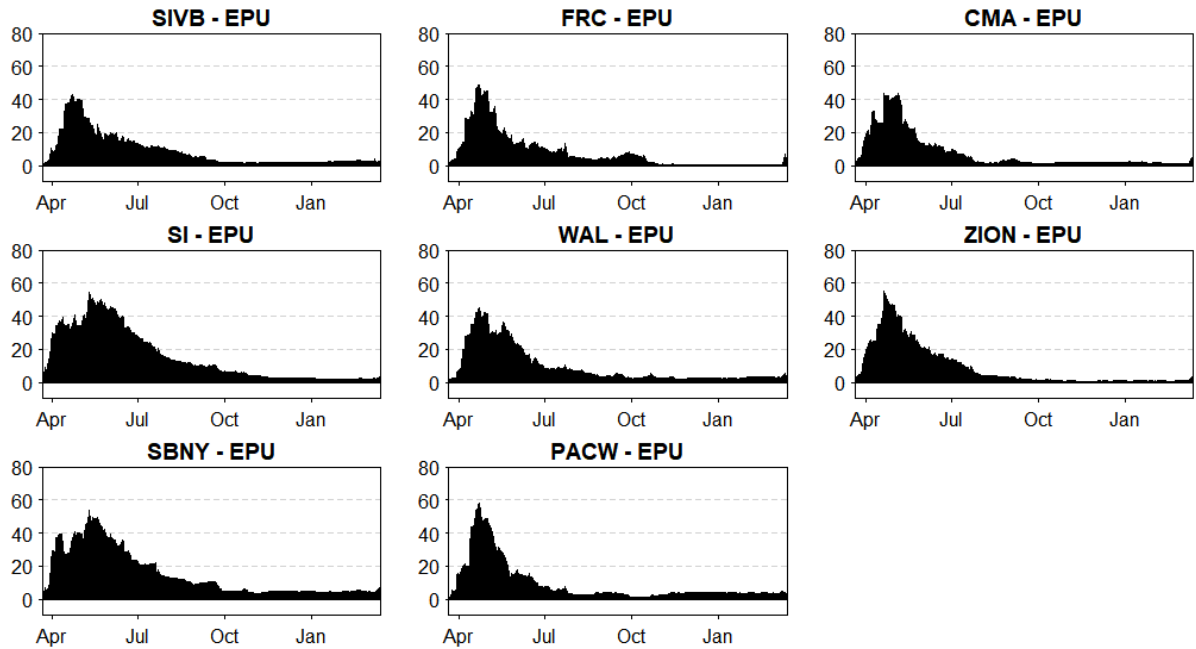

**Figure 8c.** PCI plot between EPU and bank returns. Note: The PCI values between EPU and bank returns from May 2022 to June 2022 are higher than in other periods.

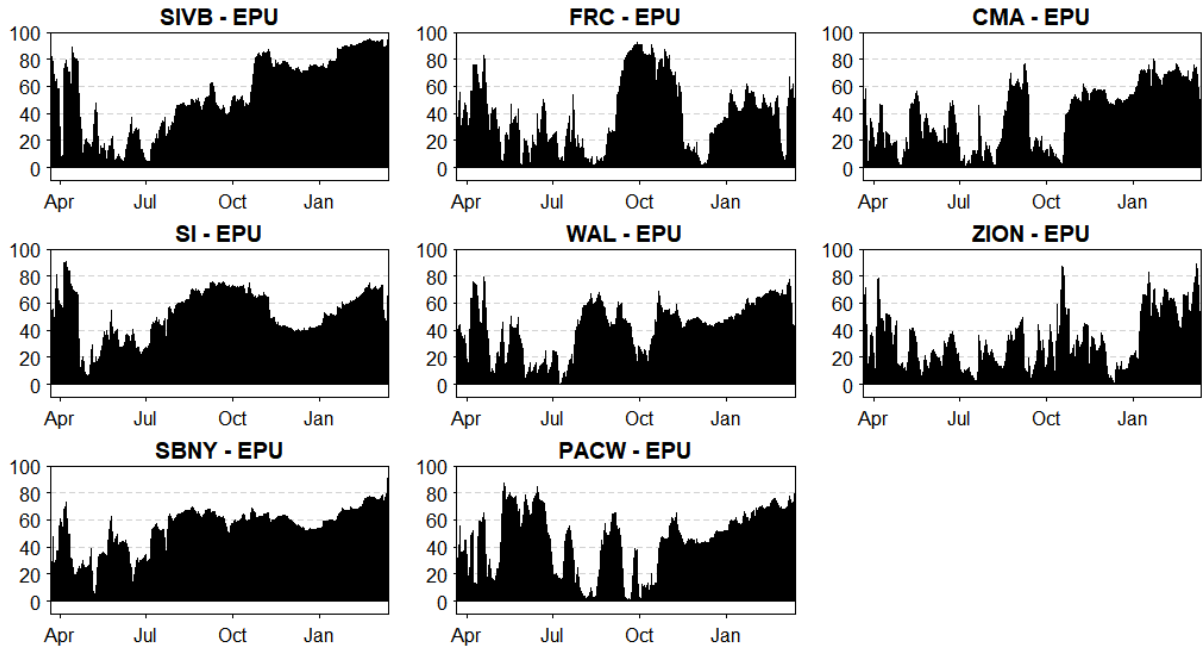

**Figure 8d.** PII plot between EPU and bank returns. Note: The PII values between EPU and bank returns of failed banks have higher values than the values between EPU and other banks.

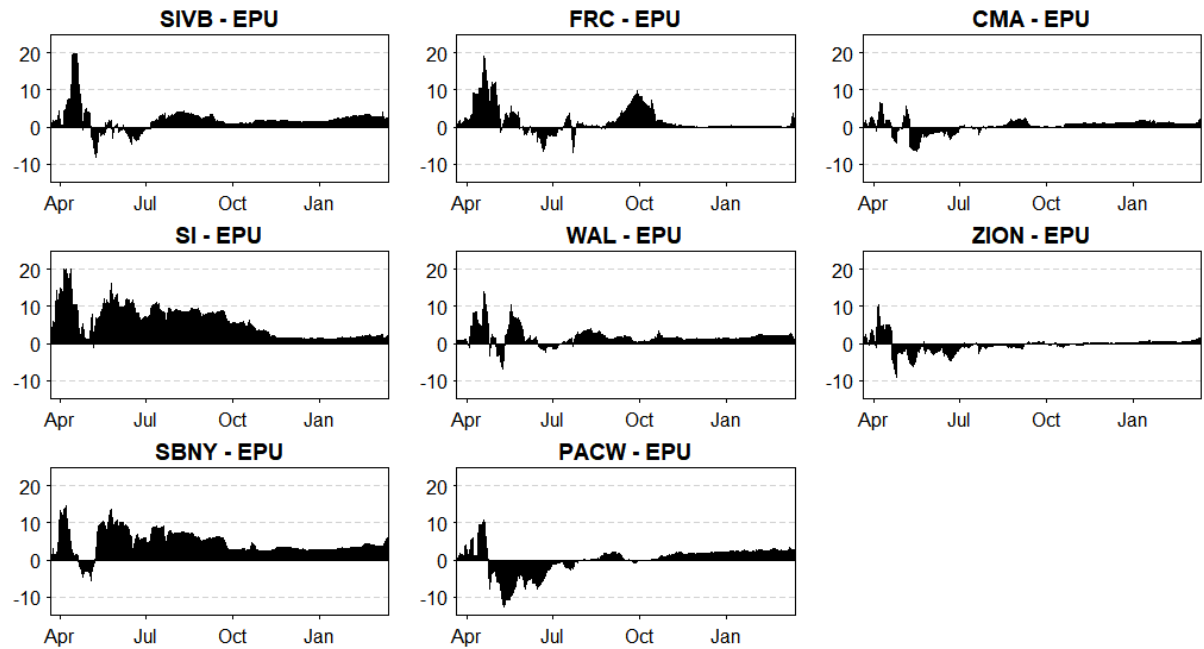

**Figure 8e.** NPDC plot between EPU and bank returns. Note: The NPDC values between EPU and failed bank returns have positive values.

## 5. Discussion

### 5.1 Impact of the economy-wide variables

The TVP-VAR connectedness analysis reveals that SOFR, SIVB, FRC, and WAL are the most significant net transmitters (givers) of shocks to banks within the financial network, with SOFR emerging as a primary driver of systemic risk propagation during early 2022. This finding complements prior research linking monetary policy to business lending and financial market volatility (Borağan Aruoba 2020). While the Federal Reserve Board uses monetary policy to balance inflation and economic activity, our dynamic analysis highlights that policy actions can have substantial spillover effects on bank stock returns, amplifying systemic vulnerability during stress episodes. SOFR's strong transmitter role before July 2022, followed by its diminished influence later in the period, suggests that the impact of monetary policy shocks on banks shifts rapidly over time.

The analysis also shows that EPU and VIX act primarily as net receivers (absorbers) of shocks. EPU, which captures economic policy uncertainty shaped by media narratives, consistently absorbs spillovers from banks and SOFR. This aligns with the finding that public perceptions of economic uncertainty are influenced by monetary policy and financial market conditions (Doms and Morin 2004). Similarly, VIX, a proxy for investor sentiment, is persistently influenced by changes in bank stock returns and SOFR, though it displays episodic spikes of positive connectedness during stress periods. These dynamics suggest that market sentiment and policy uncertainty largely respond to shocks originating from banks and monetary policy rather than acting as independent sources of systemic risk.

The episodic nature of spillovers is particularly noteworthy. Figures 5b, 6b, 7b, and 8b show that net transmitter and receiver roles fluctuate during stress episodes. SIVB and FRC, for example, acted as strong transmitters prior to October 2022, while SI, SBNY, and PACW were consistent receivers early in the period before transitioning to weak transmitters. VIX and EPU, although predominantly receivers, showed heightened connectedness with banks during the May–July 2022 stress window. This effect reinforces the importance of monitoring time-varying relationships, as systemic risk drivers can shift roles quickly and amplify contagion risk within the network.

### 5.2 Managerial implications

This paper introduces a dynamic framework for assessing systemic risk spillovers and connectedness between bank stock returns and economy-wide variables—market sentiment (VIX), monetary policy (SOFR), and economic policy uncertainty (EPU)—using TCI, NET, NPDC, PCI, and PII measures. The findings demonstrate that shocks propagate through the financial network in episodic, asymmetric patterns, with a small set of banks and SOFR acting as persistent net transmitters and the broader market variables absorbing shocks. Analyzing the risk spillovers and connectedness between bank stock returns and economic-wide variables offers several valuable insights and contributions, such as:

- Risk Management: Understanding how changes in economy-wide variables affect bank stock returns enables banks, investors, and regulators to anticipate systemic risk. Banks can adjust portfolios and capital buffers to account for shifts in monetary policy or market sentiment, while policymakers can identify early warning signals by tracking changes in bank stock return spillovers. The finding that SOFR, SIVB, and FRC amplified volatility prior to mid-2022 shows how monetary policy actions and vulnerable banks can drive systemic risk contagion.
- Macroeconomic Stability: Monitoring the interconnectedness between banks and macro variables provides actionable insights for policymakers. FRB monetary policy not only influences lending and investment activity but also shapes systemic risk transmission, particularly when banks are already vulnerable. Tracking dynamic connectedness allows policymakers to assess how effectively their actions stabilize markets and to anticipate when interventions are necessary to mitigate contagion.
- Investor Behavior: Investors' responses to changes in monetary policy, market sentiment, and economic policy uncertainty strongly influence bank stock valuations. VIX and EPU's roles as net receivers underscore how investor perceptions and policy narratives are shaped by shocks originating from the banking system and monetary policy. Recognizing these patterns helps market participants better interpret shifts in investor sentiment and adjust strategies accordingly.

Overall, this study demonstrates that bank stock returns, monetary policy, and macro-financial conditions form a highly interconnected system. Monitoring dynamic spillovers is critical for identifying vulnerabilities early, understanding shifts in systemic risk drivers, and improving the effectiveness of risk management and policy interventions.

## *5.3 Limitations and future research*

The Russell 3000 index covers over 200 banks of varied sizes and locations. A scale analysis on a more extended time series would provide more insights into the dynamic connectedness between banks and their social, legal, economic, political, and technological environments. Thus, this study presents a few limitations. First, the TVP-VAR model, limited by computer memory constraints, can only accommodate a small number of variables. Previous studies also encountered a similar limitation, typically featuring no more than five or six variables. Second, handling matrix operations for data with long time series also requires substantial computer memory, which was unavailable here. Third, other variables associated with operational level data were available, and the operational level data from SEC 10K are quarterly and annually based.

Figure 9 shows that the credit standard increases dramatically after each fiscal crisis and extraordinary event. For example, its spikes correspond with the aftermath of the 2000 Internet bubble, the 2007-2008 global financial crisis, the 2020 global COVID-19 pandemic, and the global supply chain disruption arising from the war between Russia and Ukraine. The credit squeeze in returns will make it more challenging for small businesses to access financial support and, likewise, presents challenges for underperforming financial institutions in identifying promising investment projects. Consequently, underperforming banks may take higher risks by catering to customers who could not secure loans from healthier banks. The downward spiral of credit squeezing will lead to more banks failing. The dynamic connectedness between tightening credit standards and bank failures is interesting to our future research.

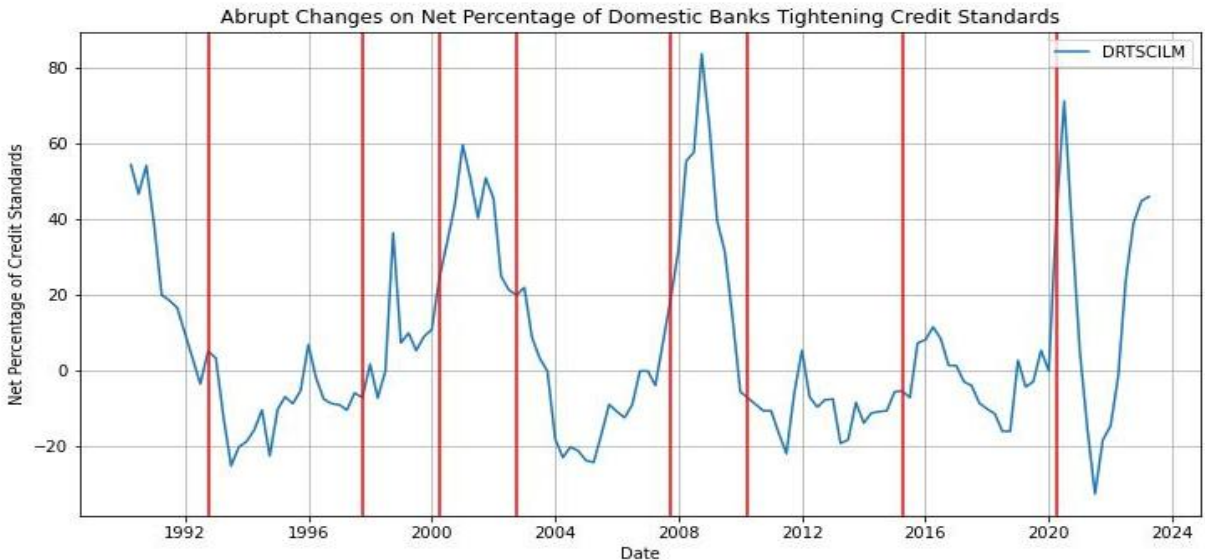

**Figure 9.** The value of DRTSCILM[5] of domestic banks from 1990 to 2023.

## 6. Conclusions

The TVP-VAR dynamic conditional connectedness analysis using rolling windows of 30 days demonstrates that a small subset of banks consistently acts as primary transmitters of systemic shocks, while others function as net receivers of spillovers. This pattern is evident not only in bank-to-bank interactions but also when economy-wide variables—market sentiment (VIX), monetary policy (SOFR), and economic policy uncertainty (EPU)—are incorporated into the network. The TCI, NPDC, PCI, and PII measures derived from the TVP-VAR model provide substantially greater explanatory power than the static connectedness measures produced by the standard VAR, revealing both the magnitude and timing of contagion pathways.

The results highlight that the identities and ranks of net transmitters and receivers remain broadly consistent across different economy-wide variables, although the magnitude of their connectedness shifts over time. SOFR, SIVB, FRC, and WAL emerge as the most prominent net transmitters, while SI, SBNY, CMA, ZION, EPU, PACW, and VIX act predominantly as net receivers. Importantly, these roles are episodic rather than static, intensifying during stress periods such as May–July 2022. This underscores the need for real-time monitoring of dynamic connectedness, as systemic risk drivers can change rapidly.

The findings confirm the presence of significant market and systemic risks within the financial network, shaped by interconnected relationships between banks and economy-wide variables. These insights are valuable for investors, policymakers, and researchers as they provide a clearer understanding of contagion dynamics and systemic vulnerabilities. Identifying dynamic transmitters and receivers enables proactive portfolio risk management, enhances early warning systems, and informs timely interventions. By quantifying how shocks propagate through the financial network and interact with key macro-financial variables, this study advances the understanding of systemic risk and offers actionable tools to mitigate its impact.

---

[5] https://fred.stlouisfed.org/series/DRTSCILM. DRTSCILM-Net Percentage of Domestic Banks Tightening Standards for Commercial and Industrial Loans to Large and Middle-Market Firms, Percent, Quarterly, Not Seasonally Adjusted

**Acknowledgment**

This work received support from the Summer Research Award at the A. R. Sanchez, Jr. School of Business and the University Research and Development Awards program at Texas A&M International University.

**Appendix A**. Variable description and integration results

**Table A1.** Assets, liabilities, and market capitals of eight American banks.

| Bank | Ticker | Assets* | Liability* | LDR(%) | Market Cap* |
|---|---|---|---|---|---|
| Silicon Valley Bank | SIVB | 211,793 | 195,498 | 42.9 | 81720.0 |
| Silvergate Capital Corporation | SI | 11,355 | 10,752 | 93.7 | 32152.6 |
| First Republic Bank | FRC | 212,639 | 195,193 | 94.1 | 33050.5 |
| Comerica Incorporated | CMA | 85,406 | 80,225 | 74.8 | 22468.3 |
| PacWest Bancorp | PACW | 42,119 | 37,278 | 84.3 | 912263.0 |
| Signature Bank | SBNY | 110,364 | 102,351 | 84.5 | 51854.5 |
| Western Alliance Bancorporation | WAL | 67,734 | 62,378 | 96.7 | 13886.4 |
| Zions Bancorporation, National Association | ZION | 89,545 | 84,652 | 77.7 | 14986.4 |

Note: *Based on December 2022 SEC 10k filing, in millions of dollars.

**Table A2.** Descriptive statistics of bank stock returns, VIX, SOFR, and EPU.

| Variable | Mean | Median | SD | Skewness | Kurtosis | J-B test | Q2(20) | 1st | 3rd |
|---|---|---|---|---|---|---|---|---|---|
| **SIVB** | -0.004 | -0.0014 | 0.033 | -5.370*** | 52.438*** | 29606.001*** | 94.495*** | -0.013 | 0.010 |
| **SI** | -0.004 | -0.0014 | 0.031 | -0.967*** | 5.775*** | 383.252*** | 106.254*** | -0.012 | 0.007 |
| **SBNY** | -0.003 | -0.0017 | 0.020 | -0.893*** | 2.686*** | 107.519*** | 28.767*** | -0.012 | 0.011 |
| **FRC** | -0.004 | -0.0021 | 0.034 | -7.634*** | 88.125*** | 82657.615*** | 10.657*** | -0.013 | 0.010 |
| **WAL** | -0.003 | -0.0010 | 0.032 | -4.683*** | 43.443*** | 20408.027*** | 21.16*** | -0.015 | 0.014 |
| **PACW** | -0.003 | -0.0012 | 0.023 | -3.428*** | 24.423*** | 6649.296*** | 20.15*** | -0.012 | 0.007 |
| **CMA** | -0.002 | -0.0014 | 0.021 | -3.135*** | 26.508*** | 7666.848*** | 107.768*** | -0.012 | 0.009 |
| **ZION** | -0.002 | -0.0008 | 0.021 | -2.365*** | 16.433*** | 3021.696*** | 129.824*** | -0.012 | 0.010 |
| **VIX** | -0.001 | -0.0081 | 0.049 | 1.141*** | 3.307*** | 166.805*** | 113.137*** | -0.030 | 0.024 |
| **SOFR** | 0.017 | 0.0000 | 0.106 | 6.207*** | 38.028*** | 16535.308*** | 159.251*** | 0.000 | 0.000 |
| **EPU** | 0 | 0.0034 | 0.039 | -0.151 | 0.441 | 2.954 | 77.18*** | -0.024 | 0.027 |

Note: $p$-value<=0.1, * $p$-value<=0.05, ** $p$-value<=0.01, and *** $p$-value<=0.005. The Ljung-Box test shows that the null hypothesis of no autocorrelation up to order 20 for the squared standard residuals (Q2(20)) cannot be rejected at the 1% significance level.

**Table A3.** ADF test results for variables and their value at first difference.

| Ticker | ADF test | $p$-value | ADF test at first difference | $p$-value |
|---|---|---|---|---|
| **SIVB** | -0.5954(0) | 0.8720 | -16.4481***(0) | 2.39E-29 |
| **SI** | -2.2075(2) | 0.2035 | -9.9326***(1) | 2.80E-17 |
| **SBNY** | -2.0992(12) | 0.2448 | -5.0298***(11) | 1.94E-05 |
| **FRC** | 0.4621(2) | 0.9837 | -8.8783***(1) | 1.34E-14 |
| **WAL** | -0.9126(1) | 0.7838 | -12.2167***(0) | 1.13E-22 |
| **PACW** | -1.7913(2) | 0.3848 | -10.4484***(1) | 1.46E-18 |
| **CMA** | -0.5924(0) | 0.8727 | -14.8754***(0) | 1.64E-27 |
| **ZION** | -0.4670(0) | 0.8983 | -9.9398***(1) | 2.68E-17 |
| **VIX** | -1.4082(9) | 0.5783 | -5.6239***(8) | 1.13E-06 |
| **SOFR** | -0.7338(0) | 0.8378 | -16.1510***(0) | 4.57E-29 |
| **EPU** | -2.7342(4) | 0.2667 | -8.4716***(10) | 1.47E-13 |

Note: $p$-value<=0.1, * $p$-value<=0.05, ** $p$-value<=0.01, and *** $p$-value<=0.005.

**Table A4.** VAR conditional correlation results.

|      | SIVB | SI   | SBNY | FRC  | WAL  | PACW | UMBF | CMA  | ZION | VIX   | SOFR  | EPU   |
|------|------|------|------|------|------|------|------|------|------|-------|-------|-------|
| SIVB | 1    | 0.42 | 0.60 | 0.56 | 0.63 | 0.77 | 0.49 | 0.51 | 0.58 | -0.41 | -0.02 | 0.02  |
| SI   |      | 1    | 0.69 | 0.34 | 0.45 | 0.44 | 0.36 | 0.40 | 0.41 | -0.58 | -0.05 | 0.11  |
| SBNY |      |      | 1    | 0.57 | 0.68 | 0.74 | 0.54 | 0.63 | 0.65 | -0.59 | -0.09 | 0.10  |
| FRC  |      |      |      | 1    | 0.80 | 0.65 | 0.63 | 0.70 | 0.74 | -0.46 | -0.04 | 0.17  |
| WAL  |      |      |      |      | 1    | 0.80 | 0.70 | 0.80 | 0.84 | -0.50 | -0.03 | 0.16  |
| PACW |      |      |      |      |      | 1    | 0.64 | 0.70 | 0.76 | -0.47 | -0.07 | 0.11  |
| CMA  |      |      |      |      |      |      | 1    | 0.88 |      | -0.52 | -0.10 | 0.16  |
| ZION |      |      |      |      |      |      |      | 1    |      | -0.56 | -0.07 | 0.14  |
| VIX  |      |      |      |      |      |      |      |      | 1    |       | 0.08  | -0.19 |
| SOFR |      |      |      |      |      |      |      |      |      | 1     |       | --0.01|
| EPU  |      |      |      |      |      |      |      |      |      |       | 1     |       |

**Table A5.** VAR partial correlation results.

|      | SIVB | SI   | SBNY | FRC   | WAL   | PACW  | UMBF  | CMA   | ZION  | VIX   | SOFR  | EPU   |
|------|------|------|------|-------|-------|-------|-------|-------|-------|-------|-------|-------|
| SIVB | 1    | 0.10 | 0.00 | 0.16  | -0.01 | 0.52  | -0.01 | -0.10 | 0.00  | -0.06 | 0.05  | -0.13 |
| SI   |      | 1    | 0.50 | -0.10 | 0.09  | -0.14 | 0.03  | 0.00  | -0.08 | -0.32 | 0.01  | 0.02  |
| SBNY |      |      | 1    | 0.04  | 0.05  | 0.34  | -0.02 | 0.06  | 0.00  | -0.13 | -0.04 | -0.05 |
| FRC  |      |      |      | 1     | 0.41  | -0.12 | 0.08  | 0.02  | 0.12  | -0.06 | 0.00  | 0.07  |
| WAL  |      |      |      |       | 1     | 0.30  | 0.10  | 0.19  | 0.16  | 0.07  | 0.11  | 0.02  |
| PACW |      |      |      |       |       | 1     | 0.07  | 0.02  | 0.13  | 0.08  | -0.06 | 0.05  |
| CMA  |      |      |      |       |       |       | 1     | 0.57  | -0.02 | -0.08 | 0.05  |       |
| ZION |      |      |      |       |       |       |       | 1     | -0.19 | 0.03  | -0.09 |       |
| VIX  |      |      |      |       |       |       |       |       | 1     | 0.03  | -0.14 |       |
| SOFR |      |      |      |       |       |       |       |       |       | 1     | -0.02 |       |
| EPU  |      |      |      |       |       |       |       |       |       |       | 1     |       |

**Table A6.** Cointegration t-test results of bank stock returns, investor sentiment index VIX, EPU, and FRB monetary policy.

|      | SIVB       | SI         | SBNY       | FRC        | WAL        | PACW       | CMA        | ZION       | VIX        | SOFR       | EPU        |
|------|------------|------------|------------|------------|------------|------------|------------|------------|------------|------------|------------|
| SIVB |            | -10.74***  | -11.68***  | -11.09***  | -11.22***  | -11.29***  | -11.57***  | -11.65***  | -10.80***  | -11.23***  | -11.17***  |
| SI   | -10.94***  |            | -10.92***  | -10.93***  | -10.04***  | -10.79***  | -10.74***  | -10.43***  | -11.73***  | -11.53***  | -11.34***  |
| SBNY | -11.52***  | -11.34***  |            | -11.60***  | -11.39***  | -11.18***  | -11.28***  | -11.46***  | -11.65***  | -11.32***  | -9.95***   |
| FRC  | -16.70***  | -8.70***   | -12.14***  |            | -12.36***  | -12.43***  | -10.63***  | -11.43***  | -9.24***   | -9.55***   | -9.55***   |
| WAL  | -9.55***   | -9.51***   | -10.32***  | -12.39***  |            | -9.99***   | -11.37***  | -10.11***  | -10.29***  | -10.78***  | -10.78***  |
| PACW | -10.27***  | -10.26***  | -13.69***  | -10.74***  | -11.14***  |            | -10.36***  | -10.37***  | -10.06***  | -10.65***  | -10.65***  |
| CMA  | -11.83***  | -11.21***  | -11.87***  | -11.51***  | -10.64***  | -11.60***  |            | -11.29***  | -11.42***  | -11.86***  | -11.79***  |
| ZION | -11.68***  | -10.45***  | -11.05***  | -10.64***  | -10.64***  | -11.43***  | -11.03***  |            | -10.79***  | -11.31***  | -11.33***  |
| VIX  | -10.95***  | -15.05***  | -10.70***  | -11.45***  | -11.00***  | -10.88***  | -11.09***  | -11.08***  |            | -11.43***  | -11.36***  |
| SOFR | -11.09***  | -11.25***  | -11.01***  | -11.04***  | -11.05***  | -11.11***  | -11.12***  | -11.05***  | -10.99***  |            | -11.08***  |
| EPU  | -13.56***  | -13.46***  | -13.64***  | -13.78***  | -13.64***  | -13.62***  | -13.72***  | -13.68***  | -13.64***  | -13.57***  |            |

Note: *p*-value<=0.1, * *p*-value<=0.05, ** *p*-value<=0.01, and *** *p*-value<=0.005.

The ADF test results in Table A3 reveal that stock returns and the economy-wide variables—market sentiment, FRB monetary policy, and EPU—are nonstationary in their original form but become stationary at the first-order difference in all samples. This suggests that these variables are integrated of order one, and therefore, the first-order difference form of the variables should be used in further analysis.

**Appendix B**. Model and formulas of static and dynamic connectedness analysis

The standard VAR model is commonly employed for static connectedness analysis and is given as:

$$x_t = d_t + \sum_{j=1}^{L} \emptyset_j x_{t-j} + w_t, \quad w_t \sim Normal(0, \Sigma_w) \quad (2)$$

In Equation (2), $x_t$ represents the vector of variables at time $t$, which includes a linear function with its own lags $L$, conformable matrices $\emptyset_j$ that represent lag dynamics, a vector of deterministic components $d_t$, and a vector of estimation errors $w_t$. In the standard VAR model, Equation (2) is valid when there is a structural stability in the parameter estimations. If structural breaks occur in the parameter estimations, then the theoretical assumption of the standard VAR is violated. Therefore, we performed a Chow test to assess the stability of the standard VAR model.

The main difference between a standard VAR and TVP-VAR model on the same shock lies in how they measure the shock's effects. The TVP-VAR model assesses the shock's effects in terms of coefficients that vary across time periods, while the standard VAR measures the shock to variable $i$ at $t$ as fixed-coefficients. The TVP-VAR reports the variance of the error as stochastic volatility (Diebold and Yilmaz 2009).

Following Cogley and Sargent (2005), the vector of stochastic volatility in the TVP-VAR model follows a geometric random walk with a normal distribution, which is here used to analyze the stochastic volatility. Return spillover and volatility connectedness analysis are captured by TVP-VAR. In addition, while the time-series data examined by VAR assumes linearity and time invariance, the TVP-VAR with stochastic volatility can manage its nonlinearity. The TVP-VAR model is then:

$$x_t = \sum_{j=1}^{p} \emptyset_t x_{t-j} + w_t, \quad w_t \sim Normal(0, S_t) \quad (3)$$

$$\emptyset_t = \emptyset_{t-1} + v_t, \quad v_t \sim Normal(0, R_t) \quad (4)$$

Where $x_t$ is a $N \times 1$ vector for $N$ variables of interest, and $p$ is the best lag length based on a Bayesian information criterion (BIC). $w_t$ is a $N \times 1$ vector, and $v_t$ is a $N^2 p \times 1$ vector. Both $w_t$ and $v_t$ are vectors of the estimated error while $S_t$ and $R_t$ are time-varying variance-covariance matrices, and $\emptyset_t$ is a matrix for the estimated coefficients of each variable at time $t$. The return spillover from bank stock $i$ to bank stock $j$ can be measured by $l_{ij}(h) = \frac{d_{ij}(h)}{\sum_{j}^{N} d_{ij}(h)}$, which is the standardized value of the element $d_{ij}(h)$ in an $h$-step ahead predicted error variance decomposition matrix in the $N \times N$ dimensional space. In this study, we report the risk spillover effect in terms of receiver, giver, the total connectedness index (TCI), the net pairwise directional connectedness (NPDC), the pairwise connectedness index (PCI), and the pairwise influence index (PII).

"Receiver" of the spillover effect is:

$$Receiver_i(h) = 100 \frac{\sum_{j=1, i \neq j}^{N} l_{ij}(h)}{\sum_{j=1}^{N} l_{ij}(h)} \quad (5)$$

"Giver" of the spillover effect is given as:

$$Giver_i(h) = 100 \frac{\sum_{j=1, i \neq j}^{N} l_{ji}(h)}{\sum_{j=1}^{N} l_{ij}(h)} \quad (6)$$

The total connectedness index of the spillover effect is given as:

$$TCI(h) = 100 \frac{\sum_{i,j=1, i \neq j}^{N} l_{ij}(h)}{\sum_{i,j=1}^{N} d_{ij}(h)} \quad (7)$$

The net pairwise directional connectedness (NPDC) is given as:

$$NPDC_{i \leftarrow j} = l_{ij} - l_{ji} \quad (8)$$

The pairwise connectedness index of spillover effect is decomposed from TCI and is given as:

$$PCI_{ij}(h) = \frac{l_{ij} + l_{ji}}{l_{ii} + l_{jj} + l_{ij} + l_{ji}} \quad (9)$$

Finally, the pairwise influence index of spillover effect is given as:

$$PII_{ij}(h) = \frac{l_{ij} - l_{ji}}{l_{ij} + l_{ji}} \quad (10)$$

The TVP-VAR calculates the dynamic conditional connectedness among banks by assessing the average impact of a shock in each bank's returns and its volatility on all other banks. The TVP-VAR reports a set of indexes (TCI) that gauge the overall interconnectedness among banks and the economy-wide variables; the $Receiver_i$ connectedness index shows the directional spillover received by variable $i$ from all other variables, while the $Giver_i$ connectedness index represents the directional spillover from variable $i$ to all other variables; the difference between the total directional $Giver_i$ and $Receiver_i$ is denoted as $NET_i$, which is the net influence of variable $i$. The NPDC measures the influence/spillover that variable $i$ has on variable $j$. If NPDC is positive, then $i$ dominates $j$, otherwise, $j$ dominates $i$.

For a large network, $NPDC_{i,j}$ is efficient computationally and more accurate compared to *TCI* in reducing bias results. The value and sign of the net total directional connectedness identify whether a specific variable is a net transmitter (giver) or a net receiver of uncertainty shocks over time and its rank in the network. The value and sign of $NPDC_{i,j}$ indicate the dynamic conditional connectedness between two specific variables. The $NPDC_{i,j}$ measure is also graphically reported in this study.

The remaining appendices are available by request from the corresponding author.

- **Appendix C**. Plots of pairwise connectedness indices (PCI) of all variables